\documentclass[preprint, longbibliography]{revtex4-1}

\usepackage[utf8]{inputenc}
\usepackage{amsmath, amssymb,graphicx}
\usepackage[cdot,mediumqspace,squaren]{SIunits}
\usepackage{braket}
\usepackage[english]{babel}

\newcommand{\Ham}{\hat{H}} 
\renewcommand{\vec}{\mathbf} 
\newcommand{\e}[1]{\text{e}^{#1}} 
\newcommand{\eps}{\varepsilon} 
\renewcommand{\k}{\vec{k}} 
\newcommand{\op}[1]{\hat{#1}} 
\renewcommand{\Re}[1]{\mathrm{Re}\left(#1\right)} 

\graphicspath{{./images/},{./}}

\begin{document}

\title{Link between the photonic and electronic topological phases in artificial graphene}

\author{Sylvain Lanneb\`{e}re\textsuperscript{1}}
\author{M\'{a}rio G. Silveirinha\textsuperscript{1,2}}
\email{To whom correspondence should be addressed:
mario.silveirinha@co.it.pt}
 \affiliation{\textsuperscript{1}
Department of Electrical Engineering, University of Coimbra and
Instituto de Telecomunica\c{c}\~{o}es, 3030-290 Coimbra, Portugal}
\affiliation{\textsuperscript{2}University of Lisbon -- Instituto
Superior T\'ecnico, Department of Electrical Engineering, 1049-001
Lisboa, Portugal}

\date{\today}

\begin{abstract}
In recent years the study of topological phases of matter has
emerged as a very exciting field of research, both in photonics and
in electronics. However, up to now the electronic and photonic
properties have been regarded as totally independent. Here, we
establish a link between the \emph{electronic} and the
\emph{photonic} topological phases of the same material system and
theoretically demonstrate that they are intimately related. We
propose a realization of the Haldane model as a patterned 2D
electron gas and determine its optical response using the Kubo
formula. It is shown that the electronic and photonic phase diagrams
of the patterned electron gas are strictly related. In particular,
the system has a trivial photonic topology when the inversion
symmetry is the prevalent broken symmetry, whereas it has a
nontrivial photonic topology for a dominant broken time-reversal
symmetry, similar to the electronic case. To confirm these
predictions, we numerically demonstrate the emergence of
topologically protected unidirectional electromagnetic edge-states
at the interface with a trivial photonic material.
\end{abstract}

\maketitle

\section{Introduction}
The discovery of topological phases of matter was a major
breakthrough in modern condensed-matter and electromagnetics
research
\cite{hasan_textitcolloquium_2010,shen_topological_2012,lu_topological_2014,lu_topological_2016,
Haldane_Nobel_2017}. The material topology is determined by the
global properties of the electronic or photonic bands, which are
characterized by some topological invariant, e.g., the Chern number.
The topological properties are robust against smooth variations of
the system parameters and can only be changed through a phase
transition that involves the exchange of topological numbers of
different bands by closing and reopening of a band-gap. This
characteristic makes the topological properties quite insensitive to
fabrication imperfections. Furthermore, perhaps the most remarkable
feature of topological materials is their ability to support
unidirectional edge states or spin-polarized edge states at the
interface with ordinary insulators
\cite{halperin_quantized_1982,hatsugai_chern_1993,kane_$z_2$_2005,raghu_analogs_2008,silveirinha_bulk-edge_2016}.
This property was demonstrated theoretically and experimentally in a
plethora of systems relying on non-reciprocal
\cite{camley_nonreciprocal_1987,zhukov_chiral_2000,haldane_possible_2008,yu_one-way_2008,wang_observation_2009,ochiai_photonic_2009,ao_one-way_2009,poo_experimental_2011,fang_realizing_2012,davoyan_theory_2013,davoyan_electrically_2014,skirlo_experimental_2015,abbasi_one-way_2015,minkov_haldane_2016,jin_topological_2016,he_photonic_2016,jin_infrared_2017}
and reciprocal materials
\cite{hafezi_robust_2011,rechtsman_photonic_2013,khanikaev_photonic_2013,gao_topological_2015,liu_gauge_2015,chen_symmetry-protected_2015,slobozhanyuk_three-dimensional_2017,silveirinha_PTD_2017}.
In particular, topological systems are quite unique platforms for
the development of integrated one-way, defect-immune electronic and
photonic guiding devices, even though other solutions not based on
topological properties may exist \cite{silveirinha_PTD_2017}.

Topological materials can be divided into two categories depending
on whether or not they remain invariant under the time-reversal
operation. Historically, the importance of the time-reversal
symmetry in the topology of physical systems was underscored by
Haldane who demonstrated in his seminal work
\cite{haldane_model_1988} that a broken time-reversal symmetry is
the key ingredient to obtain a quantized electronic Hall phase. This
important result was some decades later extended to electromagnetism
\cite{raghu_analogs_2008}, and since then a variety of strategies to
obtain nontrivial photonic topological phases with a broken
time-reversal symmetry was proposed
\cite{haldane_possible_2008,wang_observation_2009,ochiai_photonic_2009,ao_one-way_2009,poo_experimental_2011,fang_realizing_2012,skirlo_experimental_2015,minkov_haldane_2016,jin_topological_2016,jin_infrared_2017}.

In parallel, recent studies about the reflection of a light beam on
2D materials with a quantized Hall conductivity
\cite{kort-kamp_active_2015,cai_quantized_2017} have revealed
interesting connections between nontrivial electronic and photonic
topological properties (namely quantized Imbert-Fedorov,
Goos-H\"{a}nchen, and photonic spin Hall shifts).

Inspired by these ideas, here we show using the Haldane model
\cite{haldane_model_1988} how a topologically nontrivial electronic
material with a quantized Hall conductivity in the static limit can
be used as a building block to create a topologically nontrivial
photonic material. It is proven that analogous to the electronic
counterpart, the photonic band structure is topologically nontrivial
when the time-reversal symmetry is the dominant broken symmetry.
Thus, our work establishes for the first time a direct link between
the electronic and photonic topological properties.

The manuscript is organized as follows. In Sect. \ref{SectHaldane}
we propose a patterned 2D electron gas (2DEG) with the symmetries of
the Haldane model. The structure consists of an array of scattering
centers organized in a honeycomb lattice (often referred to as
``artificial graphene''
\cite{gibertini_engineering_2009,lannebere_effective_2015}) under
the influence of fluctuating static magnetic field with zero mean
value. Using a ``first-principles'' calculation method, we find the
values of the Haldane tight-binding parameters and derive the
electronic topological phase diagram. In Sect.
\ref{SectConductivity}, the dynamic conductivity response of the 2D
material is calculated with the Kubo formula. The conductivity is
used in Sect. \ref{SectConductivity} to characterize the photonic
properties of the system and derive the natural modes. The photonic
Chern numbers are found with an extension of the  theory of
\cite{silveirinha_chern_2015}. It is shown that the transition from
a trivial to a nontrivial electronic topological phase in Haldane
graphene induces a photonic topological phase transition. Thereby,
we unveil the intimate relation between the electronic and photonic
topological properties. Furthermore, it is demonstrated with
full-wave simulations that the nontrivial photonic topological phase
enables the propagation of unidirectional edge states at the
interface with an ordinary light ``insulator'' (i.e., an opaque
material with no light states). A brief overview of the main
findings of the article is presented in Sect. \ref{SecConcl}.

\section{Haldane artificial graphene}

\label{SectHaldane}

\subsection{Overview of the Haldane model}

The Haldane model is a generalization of the tight-binding
Hamiltonian of graphene to systems with a broken inversion symmetry
(IS) and/or a broken time-reversal symmetry (TRS)
\cite{haldane_model_1988}. Analogous to the graphene case, the
Haldane Hamiltonian describes a 2D hexagonal lattice with two
scattering centers per unit cell. However, in the Haldane model the
two sublattices are allowed to be different, and in the model this
feature is described by a mass term $M$, which may be positive or
negative. When the two sublattices are identical the mass term
vanishes and the 2D material is invariant under the inversion
operation. Furthermore, the Haldane model takes into account the
possible effect of a nontrivial space-varying static magnetic field
with a net flux equal to zero. The magnetic field is responsible for
breaking the TRS.

The Taylor expansion of the Haldane's Hamiltonian near $K$ and $K'$
is given by
\begin{subequations}
\begin{align}
H( \vec{K} + \vec{q}) &= -3 t_2  \cos\left( \phi \right)  \cdot \mathbf{1} + \alpha_K \cdot \boldsymbol{\sigma_z}  - \frac{3}{2} t_1 a    \left( q_x \boldsymbol{\sigma_x} +   q_y  \boldsymbol{\sigma_y}  \right), \label{E:TB_Ham_K} \\
H( \vec{K}' + \vec{q}) &= -3 t_2 \cos\left( \phi \right) \cdot \mathbf{1} + \alpha_{K'} \cdot \boldsymbol{\sigma_z} - \dfrac{3}{2} t_1 a   \left( q_x \boldsymbol{\sigma_x} -   q_y  \boldsymbol{\sigma_y}  \right),  \label{E:TB_Ham_KPrime}
\end{align}
\end{subequations}
where $a$ is the nearest-neighbors distance, $\vec{q}=q_x
\hat{\vec{x}} + q_y \hat{\vec{y}}$ is the wavevector taken
relatively to $K$ or $K'$, $t_1$ and $t_2$ are the nearest-neighbors
(in different sublattices) and next-nearest-neighbors (in the same
sublattice) hopping energies respectively, $\boldsymbol{\sigma}_i$'s
are the Pauli matrices and
\begin{subequations}
\begin{align}
\alpha_K &=M-3 \sqrt{3} t_2 \sin(\phi), \label{E:alpha_K} \\
\alpha_{K'} &= M+3 \sqrt{3} t_2 \sin(\phi), \label{E:alpha_K_Prime}
\end{align}
\end{subequations}
are the terms resulting from breaking the IS and/or the TRS at $K$
and $K'$ respectively. These parameters vanish in pristine graphene.
The phase factor $\phi$ is determined by the integral of the
magnetic vector potential along a path that joins
next-nearest-neighbors \cite{haldane_model_1988}.

The Haldane Hamiltonian leads to a two-band model whose upper and
lower band eigenfunctions, denoted by $\ket{+}$ and $\ket{-}$
respectively, have energies $E_\pm$ given by
\begin{subequations}
\begin{align}
E_\pm(K) &= - 3 t_2 \cos(\phi ) \pm \frac{1}{2} \sqrt{(3 t_1 a q)^2 +  4 \alpha_K^2 }, \label{E:Energy_K}  \\
E_\pm(K') &= - 3 t_2 \cos(\phi ) \pm \frac{1}{2} \sqrt{(3 t_1 a q)^2 +  4 \alpha_{K'}^2 }. \label{E:Energy_KPrime}
\end{align}
\end{subequations}
Of course, when the inversion and time-reversal symmetries are
preserved $\alpha_K=\alpha_{K'}=0$, and one recovers the band
diagram of pristine graphene with Dirac cones at $K$ and $K'$. On
the other hand, a nonzero $\alpha_K$ ($\alpha_{K'}$) opens an energy
gap $E_{g,K}=2|\alpha_K|$ ($E_{g,K'}=2|\alpha_{K'}|$) at $K$ ($K'$)
between the $\ket{+}$ and $\ket{-}$ bands. Remarkably the electronic
phases obtained by breaking predominantly the TRS ($
\mathrm{sgn}(\alpha_{K}) \neq \mathrm{sgn}(\alpha_{K'}) $) or the IS
($ \mathrm{sgn}(\alpha_{K}) = \mathrm{sgn}(\alpha_{K'}) $) are
topologically distinct, leading to different electronic Chern
numbers $\nu$ \cite{haldane_model_1988}. The Chern number of the
valence band $\ket{-}$ can be written as (see Appendix
\ref{sec:ChernElectronic}):
\begin{align}
\nu  = \frac{1}{2}\left[ {{\mathop{\rm sgn}} \left( {{\alpha _K}}
\right) - {\mathop{\rm sgn}} \left( {{\alpha _{K'}}} \right)}
\right]. \label{E:ElectronicChern}
\end{align}
For convenience, we will refer in the following to the electronic
phase with $\mathrm{sgn}(\alpha_{K}) \neq \mathrm{sgn}(\alpha_{K'})$
(which has a Chern number $\nu=\pm1$ and thereby a nonzero Hall
conductivity) as the ``Hall phase'' and to the electronic phase with
$ \mathrm{sgn}(\alpha_{K}) = \mathrm{sgn}(\alpha_{K'})$ (which has
vanishing Chern number $\nu=0$ and vanishing static Hall conductivity) as
the ``insulating phase''.

\subsection{Haldane model in a 2DEG}

Next, we outline how the Haldane model may be implemented by
modifying the ``artificial graphene'' structure proposed in
\cite{gibertini_engineering_2009}. The main objective is to give
some visualization of the system under study and at the same time
obtain an estimate for the Haldane's tight-binding Hamiltonian
parameters.

Artificial graphene consists of a 2DEG under the influence of a
periodic electrostatic potential $V(\vec{r})$ with the honeycomb
symmetry. As demonstrated in
\cite{gibertini_engineering_2009,lannebere_effective_2015} such
system is fully equivalent to graphene in the sense that near the
Dirac points the electrons are described by a massless Dirac
Hamiltonian with a linear energy dispersion. Therefore, by breaking
the TRS and/or the IS it should be possible to emulate the Haldane
model in this platform.
\begin{figure}[!ht]
\centering
\includegraphics[width=.95\linewidth]{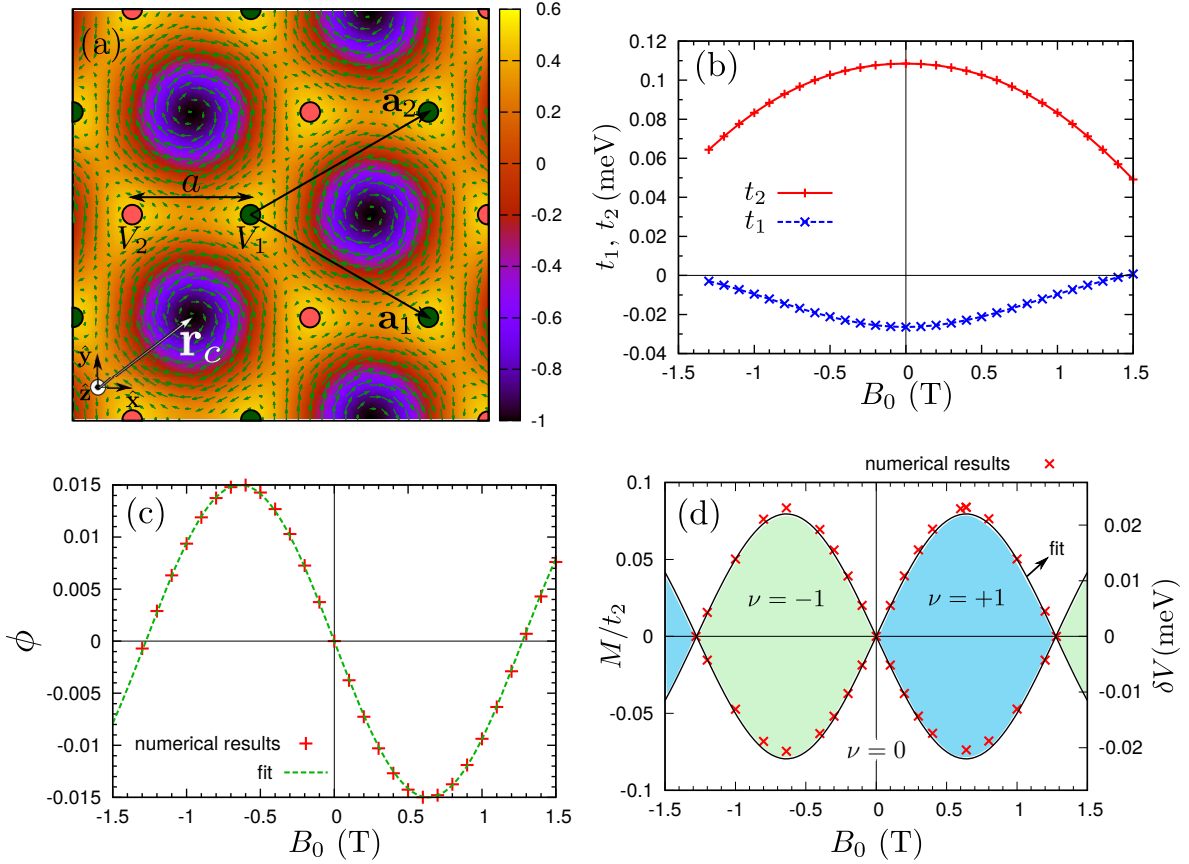}
         \caption{(a) Density plot of the magnetic field $\vec{B}$ (color scale) and of the magnetic vector potential $\vec{A}$ (arrows) used to break the TRS.
         The scattering centers (green and pink circles, associated with the electrostatic potentials $V_1$ and $V_2$, respectively),
         as well as the lattice primitive vectors are represented on the top of the fields.
         (b) Variation of the tight-binding energies $t_1$ and $t_2$ (in $\mathrm{meV}$) obtained from first-principles calculations
          as a function of the magnetic field's magnitude $B_0$ for $V_1=V_2=-0.8 ~\mathrm{meV}$.
         (c) Variation of the parameter $\phi$ obtained from a first-principles calculations as a function of $B_0$ for $V_1=V_2=-0.8 ~\mathrm{meV}$.
         (d) Phase diagram giving the electronic Chern number $\nu$ as a function of $B_0$ and $M/t_2$ (or equivalently $\delta V$) for the lower energy band.
         The boundaries between the different topological phases were obtained by interpolation of the first-principles numerical results (red crosses).}
\label{fig:FDFD_vs_Haldane_model}
\end{figure}

Following Haldane's idea, a broken IS is implemented by considering
different scattering centers for each of the sublattices of the
artificial graphene. In our model, the scattering centers are
characterized by constant potentials $V_1$ and $V_2$, and are
depicted in Fig. \ref{fig:FDFD_vs_Haldane_model}(a) as circles with
different colors. The region outside the circles has $V=0$. The
broken TRS  is achieved with a zero mean-value static magnetic field
$\bf B$. The corresponding magnetic potential $\bf{A}$, defined such
that $\vec{B}=\nabla \times \vec{A}$, is supposed to yield a
nontrivial flux $\int\limits_{}^{} {{\bf{A}} \cdot {\bf{dl}}}$ when
the starting and ending points of the integration path are
next-nearest neighbors, and a trivial flux when the starting and
ending points are nearest neighbors. As illustrated in Fig.
\ref{fig:FDFD_vs_Haldane_model}(a), a magnetic vector potential
$\vec{A}$ of the form:
\begin{equation} \label{E:vecA}
\vec{A}(\vec{r}) = \frac{3 B_0 a^2}{ 16 \pi^2 } \left[ \vec{b}_1  \sin\left( \vec{b}_1 \cdot \vec{R} \right) + \vec{b}_2 \sin\left( \vec{b}_2 \cdot \vec{R} \right) + \left( \vec{b}_1 + \vec{b}_2 \right) \sin\left( \left[ \vec{b}_1 + \vec{b}_2 \right] \cdot \vec{R} \right) \right]   \times \hat{\vec{z}}
\end{equation}
fulfills such requirements. Here $B_0$ is the peak magnetic field
amplitude in Tesla, $\vec{R}=\vec{r}-\vec{r}_c$ where $\vec{r}_c$
determines the coordinates of the honeycomb cell's center (Fig.
\ref{fig:FDFD_vs_Haldane_model}(a)) and the $\vec{b}_i$'s with
$i=1,2$ are the reciprocal lattice primitive vectors. Both $\vec{A}$
and $\vec{B}$ are represented in the honeycomb lattice in Fig.
\ref{fig:FDFD_vs_Haldane_model}(a). Note that the magnetic field is
directed along the $z$-direction, perpendicular to the 2D electron
gas. Similar to Ref. \cite{lannebere_effective_2015}, it is supposed
that the nearest-neighbors distance is $a=150\rm{nm}$ and that the
radius $R_0$ of the scattering centers satisfies $R_0/a=0.35$. Even
though challenging, in principle the required magnetic field
distribution can be created at the nanoscale by nanostructuring
permanent magnets.

The tight-binding parameters depend on $V_1$, $V_2$ and $B_0$. They
are numerically found from ``first principles'' calculations using
the effective medium formalism for electron waves developed in
\cite{silveirinha_effective_2012} and extended to artificial
graphene in \cite{lannebere_effective_2015} (we use the expression
``first principles'' in a broad sense with the meaning that the
tight-binding parameters are found from a microscopic model). The
first step of the method is to solve the Schr\"odinger equation
governing the electron wave propagation in the 2DEG, starting from
the ``microscopic'' Hamiltonian:
\begin{equation}
\Ham_\text{mic} = \frac{1}{2m_b}\left(\hat{\vec{p}} + e \vec{A}(\vec{r}) \right)^2+ V(\vec{r}),
\end{equation}
where $-e$ is the electron charge,
$V(\vec{r})=V_1(\vec{r})+V_2(\vec{r})$ is the periodic electrostatic
potential, $\vec{A}$ is given by \eqref{E:vecA} and the electron
effective mass $m_b$ is as in Ref.
\cite{gibertini_engineering_2009}: $m_b=0.067m$, with $m$ the
electron rest mass. In the spirit of Haldane's work, the spin
interaction is neglected. The Schr\"odinger equation is numerically
solved with the finite differences method
\cite{lannebere_effective_2015}. Next, following the homogenization
process detailed in \cite{lannebere_effective_2015}, we obtain a
$2\times 2$ effective Hamiltonian that determines the stationary
states and the energy dispersion. A detailed analysis (not shown) of
the energy diagrams obtained for different values of $V_1$,$V_2$ and
$B_0$ shows that the Haldane model correctly describes the physics
of the 2DEG near the Dirac points. Furthermore, the tight-binding
parameters can be calculated from a Taylor expansion of the
effective Hamiltonian near the Dirac points
\cite{lannebere_effective_2015}, which is found to be of the form
\eqref{E:TB_Ham_K}-\eqref{E:TB_Ham_KPrime}.

The numerically calculated tight-binding parameters are represented
in Fig. \ref{fig:FDFD_vs_Haldane_model}(b)-(c) as a function of the
magnetic field intensity for $V_1=V_2=-0.8 ~\mathrm{meV}$. Note that
when $V_1=V_2$ the mass parameter vanishes ($M=0$). Curiously,
unlike in graphene, in our system $|t_2|>|t_1|$. The hopping
constant $t_1$ has a value comparable with that found in
\cite{lannebere_effective_2015}. More interestingly, it can be seen
that the parameter $\phi$ obtained from the first-principles
simulations is a periodic function of $B_0$ and hence is bounded.
This feature is not present in the Haldane model wherein $\phi$ is
regarded as an arbitrary real-valued number proportional to $B_0$.
The peak value of $\phi$ found here implies that the maximum energy
gap,  $6 \sqrt{3} t_2 \sin(\phi)$, due to the applied magnetic field
is on the order of $0.156 \cdot t_2$. The peak $\phi$ is reached for
$B_0 \approx 0.64 \rm{T}$, which may be difficult to create
considering that the magnetic field varies at the nanoscale.

To characterize the topology of the 2DEG, we numerically found the
combination of parameters $B_0$ and $\delta V$ for which the band
gap closes at one of the Dirac points, taking $V_1 = - 0.8 \,
\rm{meV}$ and $V_2 = V_1 + \delta V$. A nontrivial $\delta V$
implies a nonzero tight-binding mass parameter $M$. The calculated
phase-diagram is represented in Fig.
\ref{fig:FDFD_vs_Haldane_model}(d) and shows the combination of
parameters $M$ and $B_0$ for which the band gap closes. Consistent
with \cite{haldane_model_1988}, we find that when a band gap closes
and reopens there is a topological phase transition and the
electronic Chern number changes by one unity. The Chern numbers
associated with the energy band $\left|  -  \right\rangle$ are
indicated in Fig. \ref{fig:FDFD_vs_Haldane_model}(d). The electronic
Chern number determines the static Hall conductivity in the limit of
a zero temperature when the Fermi level is in the band gap
\cite{Nagaosa, thouless_quantized_1982}. The calculated
phase-diagram agrees perfectly with Haldane's theory
\cite{haldane_model_1988}, since the periodicity of $\phi$ with
$B_0$ induces also a periodicity in the phase-diagram. Thus, similar
to \cite{haldane_model_1988}, the broken IS phase corresponds to a
trivial electronic Chern number $\nu= 0$, whereas the broken TRS
phase corresponds to a phase with $\nu\neq 0$.

As a partial summary, we outlined a physical realization of the
abstract notions developed in \cite{haldane_model_1988}, relying on
``artificial graphene" and on the periodic magnetic field potential
distribution \eqref{E:vecA}. Our study gives the tight-binding
parameters obtained from ``first-principles" calculations. In the
simulations it was assumed that $a=150 \rm{ nm}$ but the design
parameters can be renormalized to other values of $a$ through a
simple dimensional analysis (e.g., a reduction of $a$ by a factor of
2 implies an increase of all the involved energies and of $B_0$ by a
factor of 4).
Even though the practical implementation of the Haldane model in a
2DEG is admittedly very challenging, our study provides for the
first time a simple visualization of the concepts introduced in
\cite{haldane_model_1988}. In the rest of the article, we use the
tight-binding parameters obtained in this section and it is assumed
that the relation between $B_0$ and $\phi$ corresponds to the fit of
Fig. \ref{fig:FDFD_vs_Haldane_model} (c): $\phi=-0.015 \sin\left(
\frac{2\pi B_0}{2.56} \right)$, with $B_0$ in Tesla.

\section{Dynamic conductivity of Haldane graphene}
\label{SectConductivity}

In order to characterize the photonic properties of ``Haldane
graphene'', next we derive its dynamic conductivity with Kubo's
linear response theory \cite{Kubo}. It is assumed that the valence
band $\left| - \right\rangle$ is completely filled (the chemical
potential $\mu$ is in the gap) and that the temperature satisfies
$k_B T \ll E_g$, with $k_B$ the Boltzmann constant and $E_g=
2\min(|\alpha_K|,|\alpha_{K'}|)$ the gap energy. In these
conditions, the Hall conductivity in the static limit is determined
by the electronic Chern number. Furthermore, the intraband
conductivity term vanishes and thereby the dynamic conductivity is
given by \cite{mikhailov_new_2007, goncalves_introduction_2016,
Allen}
\begin{equation}
\boldsymbol{\sigma}^\text{gr}(\omega) = \frac{i e^2 \hbar}{(2\pi)^2}
\sum_{m\neq n} \int {\int { \frac{f(E_{m{\bf{k}}}) -
f(E_{n{\bf{k}}})}{(E_{m{\bf{k}}} - E_{n{\bf{k}}} - \hbar\omega
)(E_{m{\bf{k}}} - E_{n{\bf{k}}})}
\bra{n{\bf{k}}}\hat{\vec{v}}\ket{m{\bf{k}}}
\bra{m{\bf{k}}}\hat{\vec{v}}\ket{n{\bf{k}}} d^2\k } }
\end{equation}
where $f$ is the Fermi distribution function, $\hat{\vec{v}} =
\frac{1}{\hbar} \frac{\partial \Ham}{\partial \vec{k}}$ is the
velocity operator and the sum is over the different bands $\ket{+}$
and $\ket{-}$. It is implicit that the contributions of both Dirac
points are included. Somewhat lengthy but otherwise straightforward
calculations based on the continuum version of the Haldane model
show that when thermal effects are negligible ($k_B T \ll E_g$) the
dynamic conductivity is of the form
\begin{subequations} \label{E:sigma}
\begin{equation}
\boldsymbol{\sigma}^\text{gr}(\omega) = \sigma_t \vec{1}_t -
\sigma_H  \hat{\vec{z}}\times \vec{1}_t,   \label{E:sigma_tensor}
\end{equation}
where $\vec{1}_t= \hat{\vec{x}}\hat{\vec{x}} +
\hat{\vec{y}}\hat{\vec{y}}$. Thus, in general the material response
is gyrotropic, with the anti-diagonal elements of the conductivity
tensor given by $\sigma_{xy}=-\sigma_{yx} = \sigma_H$ and the
diagonal elements determined by $\sigma_{xx}=\sigma_{yy} =
\sigma_t$. For $\omega$ real valued with $\hbar|\omega| < E_g$ the
conductivity elements are given by:
\begin{equation} \label{E:sigma_t}
\sigma_t = - i \sigma_0 \frac{  \mathrm{sgn}(\omega)  }{4  } \left(
G(\Omega_K) + G(\Omega_{K'}) + \frac{ G(\Omega_{K}) - |\Omega_{K}|
}{ \Omega_{K}^2 } + \frac{ G(\Omega_{K'}) - \left| \Omega_{K'}
\right|    }{ \Omega_{K'}^2} \right),
\end{equation}
\begin{equation}\label{E:sigma_H}
\sigma_H   = \frac{ \sigma_0}{2} \left( \frac{ G(\Omega_{K'})
}{\Omega_{K'}} - \frac{ G(\Omega_{K}) }{\Omega_{K}}\right).
\end{equation}
\end{subequations}
In the above, $\Omega_i=\hbar |\omega|/(2 \alpha_i)$ is a normalized
frequency ($i=K,K'$),
$G(\Omega_i)=\tanh^{-1}\left(|\Omega_i|\right)$ and $\sigma_0=\frac{
e^2 }{h }$. Remarkably, when $k_B T / E_g \to 0$ the conductivity is
independent of the nearest neighbor hopping energy $t_1$. In the
spirit of the Haldane model, it was supposed in the conductivity
calculation that the Hamiltonian describes some particular
(non-degenerate) electron spin.

A direct inspection of Eq. \eqref{E:sigma_H} reveals that in the
absence of a magnetic field ($\phi=0$), i.e., when the parameters
$\Omega_{K}$ and $\Omega_{K'}$ are equal, the Hall conductivity is
precisely zero. In contrast, for a non-zero magnetic field two
distinct situations can occur depending on which broken symmetry is
prevalent. Indeed, for relatively low frequencies ($\hbar \omega \ll
E_g$) the function $G$ may be approximated by $G(\Omega_i) \approx
|\Omega_i|+\frac{|\Omega_i|^3}{3}$. In these conditions, the
conductivities \eqref{E:sigma_t} and \eqref{E:sigma_H} reduce to
\begin{subequations}
\label{E:QSconductivity}
\begin{align}
\sigma_t &\approx- i \sigma_0   \frac{ \omega}{\omega_0 },  \\
\sigma_H  &\approx \frac{\sigma_0   }{ 2} \left(
\mathrm{sgn}(\alpha_{K'})  -  \mathrm{sgn}(\alpha_{K})  \right) = -
 \sigma_0 \nu, \label{E:Hall_conductivity_static}
\end{align}
with $ \omega_0^{-1}=  \frac{\hbar  }{6}  \left(   \frac{  1 }{
|\alpha_{K}|  } + \frac{  1 }{ |\alpha_{K'}|  } \right)$.
\end{subequations}
Equation \eqref{E:Hall_conductivity_static} confirms that the Hall
conductivity in the static limit is quantized and is determined by
the electronic Chern number $\nu$ given by
\eqref{E:ElectronicChern}, which is the TKNN result
\cite{thouless_quantized_1982, Nagaosa}. Thus, consistent with
Haldane's work, we find that the ``insulating phase'' has a trivial
conductivity in the static limit, whereas the ``Hall phase'' has a
quantized Hall conductivity. Furthermore, near $\omega=0$ the
diagonal term $\sigma_t$ is a linear function of the frequency with
slope $\omega_0^{-1}$.
\begin{figure}[!ht]
\centering
\includegraphics[width=.95\linewidth]{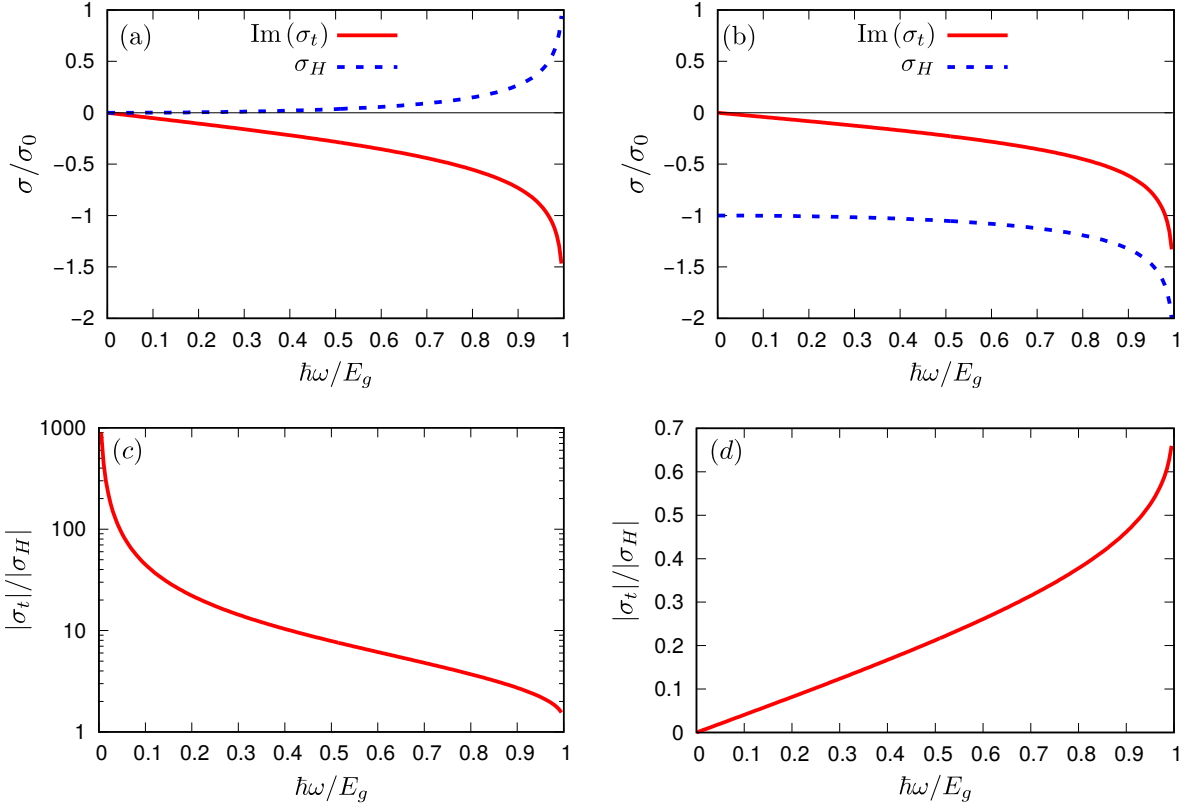}
         \caption{(a)-(b) Plot of the normalized conductivity of Haldane graphene for
 frequencies $\omega$ smaller than the gap frequency $E_g/\hbar$ in
 (a) the ``insulating'' phase: $ B_0=0.075 ~\tesla$ corresponding to $\alpha_K/\hbar \approx  9.76~\giga\hertz$ and
 $\alpha_{K'}/\hbar \approx 5.43~\giga\hertz$  and (b) the ``Hall'' phase: $B_0=0.64 ~\tesla$
corresponding to $\alpha_K/\hbar\approx 19.5  ~\giga\hertz$ and
 $\alpha_{K'}/\hbar \approx  -4.24~\giga\hertz$. (c)-(d) Ratio $|\sigma_t|/|\sigma_H|$
for the conductivities depicted in (a) and (b), respectively. In the
plots $M = 5 \cdot 10^{-3} ~\mathrm{meV}$, $t_1=-0.02 ~\mathrm{meV}$
and $t_2=0.1 ~\mathrm{meV}$.} \label{fig:conductivity_Haldane_model}
\end{figure}
The evolution of the Haldane graphene conductivity as a function of
frequency for the ``insulating'' and ``Hall'' phases is represented
in Fig. \ref{fig:conductivity_Haldane_model}. In agreement with the
phase diagram of Fig. \ref{fig:FDFD_vs_Haldane_model} (d) and with
Eq. \eqref{E:Hall_conductivity_static}, it is seen that the Hall
conductivity at $\omega = 0$ is determined by the Chern number and
is nontrivial only in the Hall phase. Perhaps the most striking
feature that discriminates the two different phases is the magnitude
ratio $|\sigma_t|/|\sigma_H|$, represented in Fig.
\ref{fig:conductivity_Haldane_model}(c) and (d). Remarkably, it is
much greater than unity for the insulating phase and near zero for
the Hall phase. The singularities in the conductivity components
near electronic band gap frequency are of logarithmic type.
\begin{figure}[!ht]
\centering
\includegraphics[width=.95\linewidth]{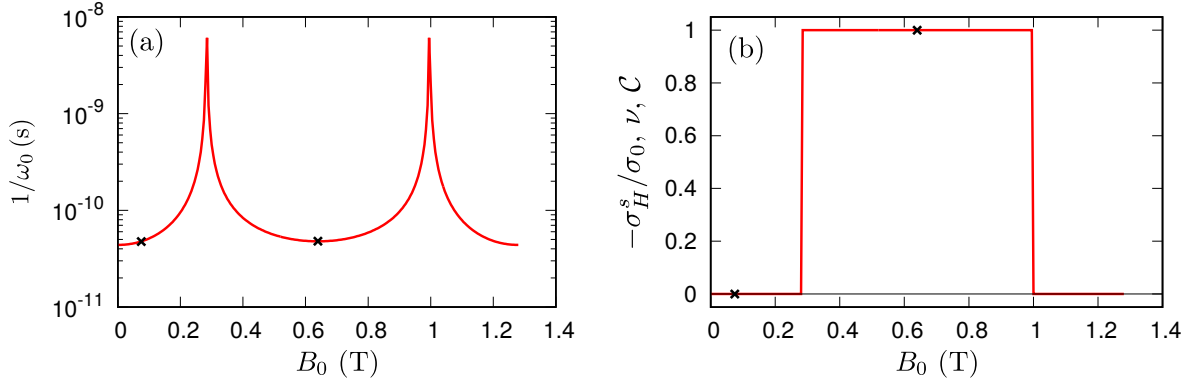}
         \caption{(a) Slope $ \omega_0^{-1}$ of $\sigma_t$
 in the quasi static limit as a function of the magnetic field $B_0$.
(b) Normalized static Hall conductivity $-\sigma_H^s$ and electronic
Chern number $\nu$
 as a function of the magnetic field $B_0$. The plot also depicts
the (positive frequency branch) photonic Chern number $\mathcal{C}$
for the structure of Fig. \ref{fig:schematic_structure}. In both
panels $t_1=-0.02 ~\mathrm{meV}$, $t_2=0.1 ~\mathrm{meV}$, $M=5
\cdot 10^{-3} ~\mathrm{meV}$. The two black crosses mark the values
of $B_0$ used in the plots of Fig.
\ref{fig:conductivity_Haldane_model}.}
\label{fig:conductivity_vs_B0}
\end{figure}

Figure \ref{fig:conductivity_vs_B0} represents the parameter
$\omega_0^{-1}$ and the static Hall conductivity $\sigma_H$ as a
function of the peak magnetic field $B_0$ for a mass parameter
$M/t_2=0.05$. By comparison with the phase diagram of Fig.
\ref{fig:FDFD_vs_Haldane_model}(d), it is seen that the Hall
conductivity $\sigma_H$ experiences discontinuous jumps at the
topological phase transitions whereas the component $\sigma_t$
remains continuous in the quasi-static limit. It is worth pointing
out that the tight-binding parameters used in Fig.
\ref{fig:conductivity_Haldane_model} (which will be adopted in the
rest of the article) yield values for $ \omega_0^{-1}$ that are
comparable for both phases.

To conclude, it is highlighted that Haldane graphene in the Hall
phase has a quasi-static conductivity response analogous to that of
a magnetized plasma. Indeed, in the limit $\omega \to 0$ a lossless
magnetized plasma is also characterized by a purely imaginary
$\sigma_t$ that vanishes in the static limit and by a nonzero (but
not quantized) $\sigma_H$ \cite{Bittencourt}. However, the two
systems are generically rather different. While a decrease of the
magnetic field amplitude $B_0$ transforms Haldane graphene into an
insulating material ($\sigma_{\omega=0}=0$), the conductivity of an
electron gas is always nontrivial even when $B_0$ is set identical
to zero because the free-electron concentration is finite. Thus, the
physical platform discussed in this article has quite unique
properties, and generically behaves differently from a 2D magnetized
plasma, particularly when the time reversal symmetry is preserved.

\section{Photonic topological properties}
\label{SectPhotonic}

Next, we characterize the photonic topological properties of the
Haldane graphene. In particular, we study the low-frequency photonic
phase transition induced by an electronic transition between the
insulating and Hall phases, and highlight the relation between the
nontrivial electronic and photonic topologies.

\begin{figure}[!ht]
\centering
\includegraphics[width=.7\linewidth]{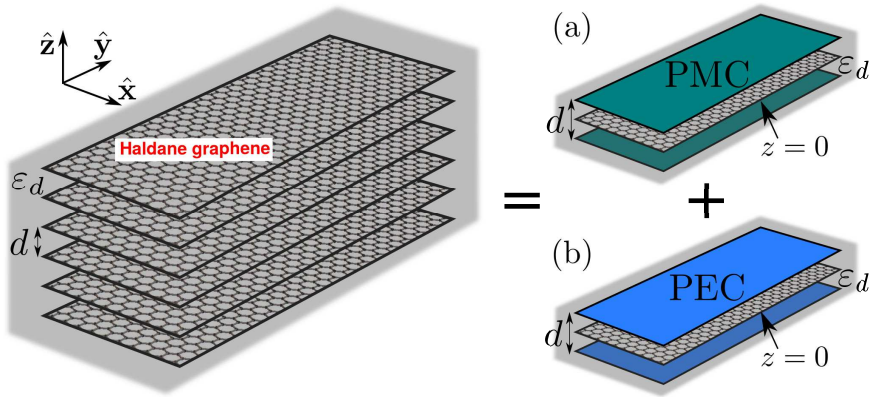}
         \caption{ The system under study: a periodic arrangement of Haldane graphene sheets with periodicity $d$ is embedded in a dielectric with
         permittivity $\eps_d$. The electromagnetic modes of this structure are the direct sum of the modes supported by a Haldane graphene sheet embedded
         in (a) a  PMC-walls waveguide and (b) a PEC-walls  waveguide, with walls located at $z=\pm d/2$.}
\label{fig:schematic_structure}
\end{figure}

The rigorous definition of Chern numbers is only possible for an
electromagnetically closed system, and thereby an isolated Haldane
graphene sheet does not provide a suitable platform to observe a
photonic topological phase transition. Thus, we consider a periodic
arrangement of Haldane graphene sheets separated by a distance $d$
and embedded in a dielectric of permittivity $\eps_d$, as shown in
Fig. \ref{fig:schematic_structure}.
We are interested in waves for which the electromagnetic energy is
allowed to flow only in the $xoy$-plane. Due to symmetry reasons the
electromagnetic modes (periodic in $z$) of the periodic system can
be split into two subsets depending on the parity of the fields with
respect to the plane $z=-d/2$: the modes of a waveguide  with
perfect magnetic conducting (PMC) walls (with $H_z$ even and $E_z$
odd) and the modes of a waveguide with perfect electric conducting
(PEC) walls (with $H_z$ odd and $E_z$ even) as illustrated in Fig.
\ref{fig:schematic_structure}.  Note that the designations even and
odd are used here with respect to a reference system with origin in
the plane $z=-d/2$.

Importantly, it can be shown that for a waveguide with PEC walls
(system of Fig. \ref{fig:schematic_structure}(b)) all the modes that
interact with the Haldane graphene are cut-off at low frequencies.
This means that to study the low frequency modes of the periodic
arrangement of Haldane graphene sheets it is enough to consider the
system of Fig. \ref{fig:schematic_structure}(a), being implicit that
the excitation should respect the indicated parity-symmetry of the
fields. For this reason in the rest of this paper we will restrict
our attention to the system with PMC walls of Fig.
\ref{fig:schematic_structure}(a).

\subsection{Natural modes}
\label{sec:naturalmodes} Next, we derive the natural (guided) modes
supported by the structure of Fig. \ref{fig:schematic_structure}(a).
Because the system is invariant to translations along the $x$ and
$y$ directions, the guided waves depend on the $x$ and $y$
coordinates as $\e{i \vec{q} \cdot \vec{r}}$ with $\vec{q}=q_x
\hat{\vec{x}} + q_y \hat{\vec{y}}$ the (transverse) wavevector.
Furthermore, in the dielectric regions the guided modes are
superposition of plane waves. The modes may be split into modes with
$H_z$ even and modes with $H_z$ odd (here, the designations even and
odd are with respect to the $z=0$ plane). The modes with $H_z$ odd
do not interact with the Haldane graphene sheet, and hence are not
interesting to us. They have a dispersion of the form $\omega
_{n{\bf{q}}}^{{\rm{odd}}} = \frac{c}{{\sqrt {{\varepsilon _d}}
}}\sqrt {{q^2} + {{\left[ {\left( {2n + 1} \right)\frac{\pi }{d}}
\right]}^2}}$, with $n=0,1,2...$, and hence are cut-off for
low-frequencies.

As to the modes with $H_z$ even, a straightforward analysis shows
that the electromagnetic field distribution that satisfies the PMC
boundary conditions ($\hat{\vec{z}}  \times \vec{H}_{z = -d/2}=
\hat{\vec{z}}  \times \vec{H}_{z =d/2}=0$) and ensures the
continuity of the tangential electric field at the
graphene-dielectric interface is of the form:
\begin{subequations}
\begin{align}
  \vec{H}(\vec{r}) & = \e{i \vec{q} \cdot \vec{r}} \left[ \mathrm{sgn}(z)\sinh\left( \gamma_0 \left[ \frac{d}{2} - |z|  \right]  \right) \left(C_1   \vec{q}  \times \hat{\vec{z}} + C_2\vec{q} \right)  + i  \frac{q^2}{\gamma_0} C_2 \cosh\left( \gamma_0 \left[ \frac{d}{2} - |z|  \right]  \right)   \hat{\vec{z}} \right] \label{E:vecH_PMC}\\
  \vec{E}(\vec{r}) &= \frac{i \e{i \vec{q} \cdot \vec{r} } }{\omega \eps_0 \eps_d } \left[ \cosh\left( \gamma_0 \left[ \frac{d}{2} - |z|  \right]  \right) \left(  -  \gamma_0    C_1  \vec{q} + C_2   \left(  \gamma_0 - \frac{ q^2 }{\gamma_0  } \right) \vec{q} \times \hat{\vec{z}} \right)  \right. \label{E:vecE_PMC} \\
  & \hspace{7.5cm} \left. -   i  C_1~\mathrm{sgn}(z)  q ^2 \sinh\left( \gamma_0 \left[ \frac{d}{2} - |z|  \right]  \right)  \hat{\vec{z}}  \right], \nonumber
\end{align}
\end{subequations}
where $\gamma_0=\sqrt{q^2- \eps_d \omega^2/c^2 }$,
$q=\sqrt{q_x^2+q_y^2}$, and $C_1$, $C_2$ are (unknown)
complex-valued coefficients. Using the boundary condition for the
tangential component of the magnetic field at the
graphene-dielectric interface, $\hat{\vec{z}} \times \left(
\vec{H}_{z = 0^+} - \vec{H}_{z = 0^-} \right)=
\boldsymbol{\sigma}^{\rm{gr}}  \cdot \vec{1}_t \cdot \vec{E}$, one
obtains the following homogeneous system of equations
 \begin{align}  \label{E:characteristic_matrix_equation}
 \begin{pmatrix}  \frac{2 i \omega \eps_0\eps_d }{\gamma_0} -  \sigma_t \mathrm{coth}\left( \gamma_0 \frac{d}{2}\right)   & - \sigma_H  \mathrm{coth}\left( \gamma_0 \frac{d}{2}\right)  \left(  \frac{\gamma_0^2- q^2 }{\gamma_0^2  }\right) \\ \sigma_H \mathrm{coth}\left( \gamma_0 \frac{d}{2}\right) \left(   \frac{\gamma_0^2  } {\gamma_0^2- q^2 }\right)  &  -\sigma_t \mathrm{coth}\left( \gamma_0 \frac{d}{2}\right)  +     \frac{-2   i \gamma_0  } {\omega \mu_0 } \end{pmatrix} \cdot \begin{pmatrix}  C_1  \\  C_2  \end{pmatrix} = 0,
\end{align}
whose solutions give the natural modes of oscillation of the system.
The dispersion equation is obtained by setting the determinant of
the matrix identical to zero
 \begin{align} \label{E:dispersion_plasmons_Haldane_graphene}
 \left( 2 i \frac{\omega \eps_0\eps_d }{\gamma_0} -  \sigma_t \mathrm{coth}\left( \gamma_0 \frac{d}{2}\right) \right) \left( -\sigma_t \mathrm{coth}\left( \gamma_0 \frac{d}{2}\right)  + 2   i    \frac{\gamma_0  } {-\omega \mu_0 }\right) +  \sigma_H^2   \mathrm{coth}^2\left( \gamma_0 \frac{d}{2}\right)     = 0.
\end{align}
The solutions  $\omega=\omega_{n\vec{q}}$ of the above equation
determine the photonic band diagram. It is interesting to note that
the limit $\gamma_0 d \to \infty$ yields the standard dispersion
equation of magnetoplasmons
\cite{chiu_plasma_1974,ferreira_confined_2012,goncalves_introduction_2016}
(with no PMC walls). Furthermore, a similar analysis shows that the
dispersion of a waveguide with PEC walls is given by a similar
expression with ``$\tanh$'' in the place of ``$\coth$''.

The electromagnetic modes $ \vec{E}_{n\vec{q}}$, $
\vec{H}_{n\vec{q}}$ associated with a given $\omega_{n\vec{q}}$ are
obtained from Eqs. \eqref{E:vecH_PMC} and \eqref{E:vecE_PMC} with
the coefficients $C_1$ and $C_2$ given by (see Eq.
\eqref{E:characteristic_matrix_equation})
\begin{subequations}
 \begin{align} \label{E:coeff_eigenvector_C1}
 C_1 &=   \sigma_H  \mathrm{coth}\left( \gamma_{0,n\vec{q}} \frac{d}{2}\right)  \left(  \frac{\gamma_{0,n\vec{q}}^2- q^2 }{\gamma_{0,n\vec{q}}^2  }\right), \\
 C_2 &= 2 i \frac{\omega_{n\vec{q}} \eps_0\eps_d }{\gamma_{0,n\vec{q}}} -  \sigma_t \mathrm{coth}\left( \gamma_{0,n\vec{q}} \frac{d}{2}\right),\label{E:coeff_eigenvector_C2}
\end{align}
\end{subequations}
and $\gamma_{0,n\vec{q}}=\sqrt{q^2- \eps_d
\omega_{n\vec{q}}^2/c^2}$.

It is useful to consider the solutions of the dispersion equation
\eqref{E:dispersion_plasmons_Haldane_graphene} for a conductivity
model with $\sigma_t=0$ and a constant (frequency independent)
$\sigma_H$. For $\sigma_H=\pm \sigma_0$ this model reduces to the
static conductivity of Haldane graphene in the ``Hall phase''. As
shown below, it provides a good approximation of the optical
response of Haldane graphene at low frequencies. The modal
dispersion with this model is
\begin{align} \label{E:disperion_mode_analytic}
 \omega _{n{\bf{q}}}^{{\rm{even}}} \approx \frac{c}{\sqrt{\eps_d }}\sqrt{ q^2 + \left( \frac{2}{d} \arctan \left(  \sqrt{\frac{\mu_0}{\eps_0\eps_d }} \frac{\sigma_H}{2}   \right) + \frac{2 \pi n}{d} \right)^2
 },
\end{align}
with $n=0,1,...$. The modes with $n \ge 1$ are evidently cut-off for
low frequencies, and hence in the following we focus on the mode
with $n=0$.
In the limit $\sigma_H=0$ this mode follows the light line, and is
clearly a transverse electromagnetic (TEM) wave with magnetic field
along $z$ and electric field parallel to the plates. For a finite
$\sigma_H$, the $n=0$ mode interacts with the Haldane graphene sheet
and this opens a band gap in the dispersion diagram, with a cut-off
frequency $\omega_\text{BG}  \equiv \frac{2c}{d\sqrt{ \eps_d  }}
\arctan \left( \sqrt{\frac{\mu_0}{\eps_0\eps_d }} \frac{\sigma_H}{2}
\right)$. The cut-off frequency is inversely proportional to the
distance between the waveguide walls and approaches zero when
$\sigma_H\to0$. The attenuation factor in the direction normal to
the graphene plane is $\gamma_{0,\vec{q}}=\frac{2i}{d} \arctan
\left( \sqrt{\frac{\mu_0}{\eps_0\eps_d }}
\frac{\sigma_H}{2}\right)$, and hence it is pure imaginary for
$\sigma_H\neq0$ implying that the natural mode is not guided by the
2D material but rather by the waveguide walls.

In the case of a quantized Hall conductivity $\sigma_H=\sigma_0$,
the cut-off frequency can be expressed in terms of the fine
structure constant $\alpha$: $\omega_\text{BG}  =  \frac{2c}{d\sqrt{
\eps_d }} \arctan \left(  \frac{ \alpha }{ \sqrt{\eps_d }} \right)$
and the associated wavelength (in host dielectric) at the gap edge
is well approximated by $\lambda_\text{gap} \approx \frac{ \pi
\sqrt{ \eps_d } d  }{ \alpha} \approx  430.5 \sqrt{ \eps_d }  d$.
Thus, for this small value of $\sigma_H$ the distance $d$ is
ultra-subwavelength at the cut-off frequency.

\subsection{Band diagrams and photonic Chern numbers}

\begin{figure}[!ht]
\centering
\includegraphics[width=.95\linewidth]{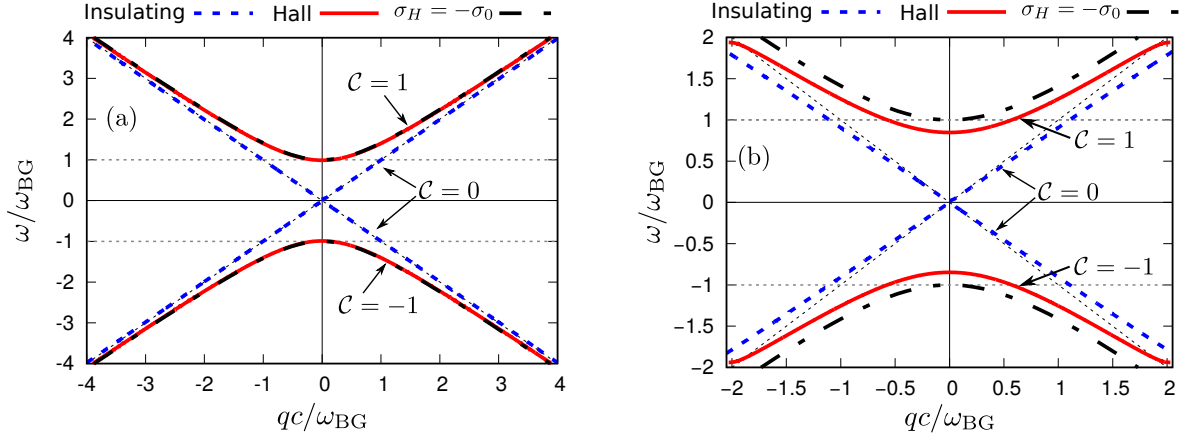}
         \caption{ Dispersion diagram of the lowest (positive and negative) frequency modes for $\eps_d=1$ and
         (a) $d=2~\centi\meter$ corresponding to $\omega_\text{BG} \approx 0.22 ~\giga\hertz$
         (b) $d=1~\milli\meter$ corresponding to $\omega_\text{BG} \approx 4.38 ~\giga\hertz $.
         The curves associated with the insulating phase were obtained with the conductivity of Fig. \ref{fig:conductivity_Haldane_model}(a)
         and the curves associated with the Hall phase with the conductivity of Fig. \ref{fig:conductivity_Haldane_model}(b).
         The curves labeled by $\sigma_H=-\sigma_0$ correspond to the analytical solution \eqref{E:disperion_mode_analytic} for the
         simplified conductivity model $\sigma_t=0$ and $\sigma_H=-\sigma_0$.
         The photonic Chern number $\mathcal{C}$ for each dispersion branch is given in the insets.}
\label{fig:band_diagram_plasmons}
\end{figure}

Figure \ref{fig:band_diagram_plasmons} shows the low-frequency
photonic band diagram for the different electronic topological
phases of Haldane graphene and two values of the distance $d$. In
this calculation, we used the conductivity responses of Fig.
\ref{fig:conductivity_Haldane_model}. Note that the photonic band
diagram shows both positive and the negative frequency solutions.

Consistent with the discussion in Sect. \ref{sec:naturalmodes}, the
band diagrams reveal (for both values of $d$) that the dispersions
induced by the distinct electronic phases are different: in the
insulating phase the dispersions follow closely the light line,
whereas in the Hall phase the diagram has a low-frequency band gap.
The modes in the insulating phase lie outside the light cone and
hence are guided by the graphene sheet. In contrast, in the Hall
phase the dispersion lies inside the light cone and the wave is --
as predicted by the static conductivity model-- guided by the
waveguide walls. Furthermore, as seen in Fig.
\ref{fig:conductivity_Haldane_model} the static conductivity model
with $\sigma_t=0$ and $\sigma_H=-\sigma_0$ gives overall a fairly
good approximation of the modal dispersion. The approximation is
better for larger values of $d$, when $\omega_\text{BG} \ll
\omega_\text{max}$ with $\omega_\text{max} = E_g/\hbar$ the
interband absorption threshold.

From the eigenmodes expression \eqref{E:vecH_PMC}  and
\eqref{E:vecE_PMC} it is possible  to compute the Chern number for
each photonic band. The calculation details are given in Appendix
\ref{sec:Berry_potential_layered_media}. The approach is based on an
extension to layered structures of the theory of
\cite{silveirinha_chern_2015, silveirinha_bulk-edge_2016}.

The values of the photonic Chern number for each branch  are
depicted as insets in Fig. \ref{fig:band_diagram_plasmons}. The
photonic Chern numbers of the bands associated with the electronic
insulating phase are equal to zero, whereas they are nontrivial,
with values $\mathcal{C}=\pm 1$, for the bands associated with the
Hall phase. Thus, the electronic topological transition from the
insulating to the Hall phase (see Fig.
\ref{fig:conductivity_vs_B0}), induces the band-gap opening at
$\omega=0$ accompanied by the exchange of photonic Chern numbers
between the positive and negative frequency bands, and thereby a
photonic topological phase transition. As expected
\cite{avron_homotopy_1983}, the total Chern number for each phase is
conserved throughout this transition. Remarkably, the topological
photonic properties of the material are directly linked to its
electronic counterparts such that the electronic Chern number of the
valence band is equal to the photonic Chern number of the positive
frequency branch: $\mathcal{C}=\nu=-\sigma_H^s/\sigma_0$, as shown
in Fig. \ref{fig:conductivity_vs_B0}(b).
This result proves that a biasing magnetic field with a vanishing
flux enables a nontrivial photonic topology, similar to the Haldane
result for the electronic case.

\subsection{Unidirectional edge-states}

According to the bulk-edge correspondence
\cite{silveirinha_bulk-edge_2016}, one may expect that an interface
of the Hall phase of Haldane graphene and a trivial photonic
insulator may support topologically protected edge-states that span
the entire common band-gap. The existence of such unidirectional
topologically protected edge-states is demonstrated next with full
wave simulations. The trivial photonic insulator is implemented with
the same waveguide but with a PEC plate in the place of the Haldane
graphene. It may be checked that the modes supported by such a
structure
are  cut-off in the long wavelength limit.

We used CST Microwave Studio \cite{CST} to demonstrate the emergence
of the topological edge states. The optical response of a 2D
material sheet with 2D conductivity $\boldsymbol{\sigma}^\text{gr}$
can be emulated with an equivalent 3D material with thickness $h_0$
and an equivalent permittivity $ \boldsymbol{\eps}/\eps_0 =
\boldsymbol{1}+i\frac{\boldsymbol{\sigma}^\text{gr}}{h_0 \omega
\eps_0}$. The thickness $h_0$ must be much smaller than the
wavelengths of interest and in addition $h_0 \ll d$, and was taken
equal to $h_0=d/10$ in the numerical study.

In the simulations, the response of the Haldane graphene
conductivity in the Hall phase is as shown in Fig.
\ref{fig:conductivity_Haldane_model}(b). The waveguide height is $d
= 400 ~\micro\meter$. The structure is excited with a small dipole
antenna (polarized along the $x$-direction) located in the close
proximity of the interface of the two waveguide regions, with the
oscillation frequency in the band-gap ($\omega < \omega_\text{BG}$).
\begin{figure}[!ht]
\centering
\includegraphics[width=.95\linewidth]{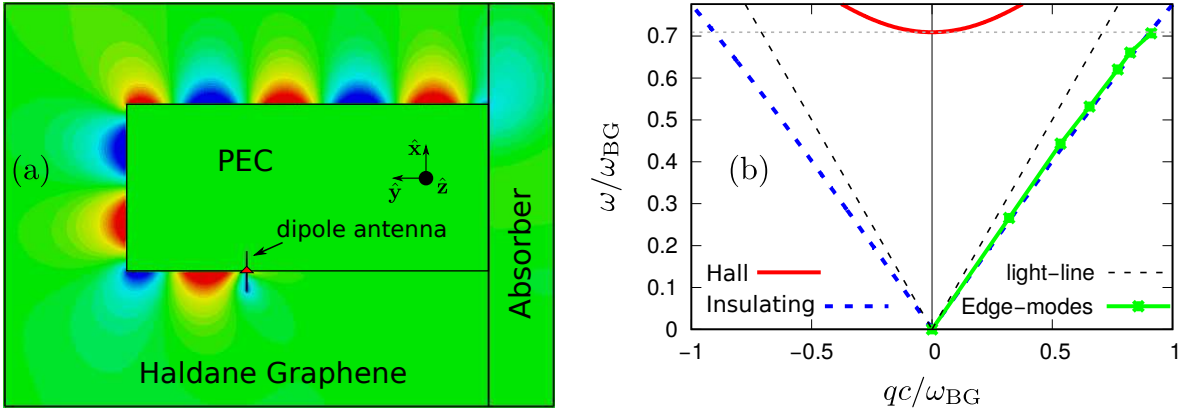}
         \caption{(a) Time snapshot of the magnetic field $H_z$ radiated
         by a short dipole antenna with $\omega=  0.62~\omega_\text{BG}$ showing the excitation of the topological edge state.
         (b) Dispersion of the unidirectional edge-modes supported by the system (discrete points joined by a solid line obtained by
         interpolation). The band structure of the bulk material in
         the Hall and insulating phases is also represented in the
         figure.
         In the plots the dielectric host is air ($\eps_d=1$) and the distance between the waveguide plates is $d = 400 ~\micro\meter$.
         }
\label{fig:edge_state}
\end{figure}
A time snapshot of the $z$ component of the magnetic field is
represented in Fig. \ref{fig:edge_state}(a) for the frequency
$\omega \approx 0.62 \omega_\text{BG}$. As seen, a unidirectional
edge state propagating along the $+y$ direction is excited in the
bottom region at the interface between the PEC material and Haldane
graphene. A time animation of the magnetic field is available in the
supplementary online materials \cite{Sup_material} and further
highlights that this edge state propagates along the interface,
regardless of the sharp corners, before reaching an absorber on the
right side region.
By fitting the wavelength of the guided modes we numerically
determined the dispersion of the edge-states as presented in Fig.
\ref{fig:edge_state}(b). As seen, the edge-modes exist in the entire
band gap and closely follow the dispersion of the insulating
photonic phase of the Haldane graphene. The low frequency part of
the curve was obtained by interpolation because the guided
wavelength approaches infinity.

In the waveguide environment, the edge-state is topologically
protected against the scattering by an arbitrary three-dimensional
defect, and in particular it is protected against sideway scattering
\cite{gangaraj_effects_2016}. The same property holds (for the
relevant wave polarization, i.e., for quasi-transverse magnetic
waves) in a periodic array of Haldane graphene sheets, but in this
case only for two-dimensional defects uniform along the
$z$-direction. Indeed, for a periodic material there are additional
radiation channels, for example, the modes that propagate off-plane
or the TEM wave with electric field normal to the graphene plane.
Note that this constraint on the defects geometry and wave
polarization also applies to conventional designs based on
gyrotropic media \cite{gangaraj_effects_2016}.

Even though in the numerical simulations the response of the Haldane
graphene was assumed lossless (consistent with the theoretical
model), we checked that the presence of moderate loss does not
affect significantly the edge state propagation in the simulation
(not shown).

\section{Conclusions}
\label{SecConcl}

The work developed in this article is divided into two parts. In the
first part we verified using a ``first-principles'' mean field
theory the validity of the Haldane model for a honeycomb lattice
with broken TRS and IS. We proposed a magnetic field distribution
that mimics the main features of Haldane's theory, and found the
dependence of the tight binding parameters on the magnetic field in
artificial graphene. The electronic phase diagram showing the range
of parameters for which the magnetized artificial graphene is
topologically nontrivial and the quantized Hall conductivity is
nonzero was determined.

In the second part, we investigated the optical response of the 2D
topological material. The dynamic electric conductivity of Haldane
graphene was found with the Kubo formula for a filled valence band.
Using this result, we determined the guided modes and the
(low-frequency) photonic band diagram of a periodic stack of Haldane
graphene sheets. As a fingerprint of the quantized conductivity, the
cut-off frequency of the low-frequency band-gap is written in terms
of the fine structure constant.

Furthermore, our analysis reveals that the electronic phase
transition between the insulating and Hall phases induces a photonic
phase transition through the opening of a band-gap at $\omega=0$ and
an exchange of Chern numbers between the positive and negative
frequency bands.  Interestingly, the electronic and photonic
topological properties of this system are intimately related, and we
find that the electronic and photonic Chern numbers are identical.
In particular, our results imply that a biasing magnetic field with
zero net flux can induce a nontrivial topological photonic response.
Finally, in agreement with the bulk-edge correspondence, it was
shown that the nontrivial photonic phase of Haldane graphene
supports topologically protected unidirectional edge states at the
interface with an ordinary photonic insulator.

\appendix

\section{The electronic Chern number} \label{sec:ChernElectronic}

In the framework of the exact Haldane Hamiltonian, the electronic
Chern number of the $n$-th band is given by
\begin{align}
{\nu _n} = \frac{1}{{2\pi }}\int\limits_{{\rm{BZ}}} {\int
{{d^2}{\bf{k}}\;{\bf{\hat z}} \cdot \nabla  \times
{\vec{\mathcal{A}}_{n{\bf{k}}}}} },
\end{align}
where ${\vec{\mathcal{A}}_{n{\bf{k}}}}$ is the Berry potential and
the integration is over the entire Brillouin zone. The Berry
potential is written in terms of the energy eigenfunctions. A
suitable globally defined gauge of eigenfunctions for the valence
band $\ket{-}$ is
\begin{align}
\psi _{\bf{k}}  = \left( {\begin{array}{*{20}{c}}
{ - {h_{12}}\left( {\bf{k}} \right)}\\
{{h_{11}}\left( {\bf{k}} \right) - {E_ - }\left( {\bf{k}} \right)}
\end{array}} \right).
\end{align}
In the above, ${{h_{ij}}}$ with $i,j=1,2$ represent the elements of
the exact Haldane Hamiltonian and $E_-$ is the exact energy
dispersion of the valence band (see Eq. (1) of Ref.
\cite{haldane_model_1988}). The Berry potential of the valence band
is given by:
\begin{align}
{\vec{\mathcal{A}}_{n{\bf{k}}}} = \frac{{{\mathop{\rm Re}\nolimits}
\left\{ {i\psi _{\bf{k}}^{*} \cdot {\partial _{\bf{k}}}\psi
_{\bf{k}} } \right\}}}{{\psi _{\bf{k}}^{*} \cdot  \psi _{\bf{k}}}}.
\end{align}
It is evidently a smooth function of the wave vector, except
possibly at the points wherein the considered gauge vanishes, i.e.,
at the points of the Brillouin zone for which ${{h_{12}}\left(
{\bf{k}} \right)}=0$ and ${{h_{11}}\left( {\bf{k}} \right) = {E_ -
}\left( {\bf{k}} \right)}$. A straightforward analysis using the
analytical formula of ${{h_{12}}}$ \cite{haldane_model_1988},
reveals that the only possible singularities are the high-symmetry
points $K$ and $K'$. Hence, using Stokes theorem it is possible to
reduce the calculation of the valence band Chern number to two
contour integrals surrounding the points $K$ and $K'$:
\begin{align} \label{Ap:ElecChern}
\nu =  - \frac{1}{{2\pi }}\sum\limits_{i = K,K'} {}
\oint\limits_{{C_i}} {{{\vec{\mathcal{A}}}_{\bf{k}}} \cdot
{\bf{dl}}}.
\end{align}
Here, $C_i$ stands for a circumference (with anti-clockwise
orientation) of arbitrarily small radius centered at the point
$i=K,K'$. Explicit calculations show that:
\begin{align}
 \frac{1}{{2\pi }}
\oint\limits_{{C_i}} {{{\vec{\mathcal{A}}}_{\bf{k}}} \cdot
{\bf{dl}}} =
 s_i {\frac{1}{2}\left( {1 - {\mathop{\rm sgn}} \left( {{\alpha _i}}
\right)} \right)}
\end{align}
with $s_K = 1$ and $s_{K'} = -1$. Substitution of this result into
Eq. \eqref{Ap:ElecChern} yields the electronic Chern number
\eqref{E:ElectronicChern}.

\section{The photonic Chern number} \label{sec:Berry_potential_layered_media}

In this appendix, we present the derivation of the Berry potential
and photonic Chern number for the system of Fig.
\ref{fig:schematic_structure}. The derivation is an extension of the
theory developed in \cite{silveirinha_chern_2015, SilvDisp} to the
case of $z$-stratified inhomogeneous closed systems. For the sake of
brevity, we reuse here the notations and concepts introduced in
\cite{silveirinha_chern_2015}. For more information the reader is
referred to \cite{silveirinha_chern_2015}.

The electromagnetic modes of the system satisfy the homogeneous
Maxwell's equations
\begin{align} \label{E:Maxwell_equation_freq}
  \op{N} \cdot \vec{f} = \omega \vec{M} \cdot \vec{f},
\end{align}
where $\vec{f}=\left(\vec{E} \quad \vec{H} \right)^T$ is a
six-vector whose components are the electric and the magnetic
fields. In the above, $\op{N}$ is a differential operator
\begin{align}
\op{N}= \begin{pmatrix} 0 & i \nabla \times \vec{1}_{3\times3} \\ -i
\nabla \times \vec{1}_{3\times3} & 0 \end{pmatrix},
\end{align}
and $\vec{M}$ is the material matrix. For the system of Fig.
\ref{fig:schematic_structure} with a conductivity sheet centered at
$z=0$, the material matrix is of the form
\begin{align}
\vec{M}(\vec{r})
 &= \vec{M}_d + \delta(z) \begin{pmatrix} \dfrac{\boldsymbol{\sigma}}{-i \omega}  & 0 \\ 0 & 0 \end{pmatrix}
\end{align}
where $\boldsymbol{\sigma}$ is the 2D conductivity and $\vec{M}_d
=\begin{pmatrix} \eps_0\eps_d \vec{1}_{3\times3}  & 0 \\ 0 & \mu_0
\vec{1}_{3\times3} \end{pmatrix}$ is the material matrix of the
surrounding dielectric.

The electromagnetic fields are Bloch waves, $\vec{f}(\vec{r})=
\vec{f}_{n\vec{q}}(z) \e{i \vec{q}\cdot\vec{r}}$, where the field
envelope $\vec{f}_{n\vec{q}}$ depends only on the $z$ coordinate and
$\vec{q}=q_x \vec{\hat x}+q_y \vec{\hat y}$. The Berry potential
$\vec{\mathcal{A}}_{n\vec{q}}$ is defined from the Hermitian
formulation of the Maxwell equations, and can be written as
\begin{equation}\label{E:def_Berry_potential}
 \vec{\mathcal{A}}_{n\vec{q}}= \frac{i \braket{\vec{Q}_{n\vec{q}}|\partial_\vec{q} \vec{Q}_{n\vec{q}}}}{\braket{\vec{Q}_{n\vec{q}}| \vec{Q}_{n\vec{q}}}},
\end{equation}
where the $\vec{Q}_{n\vec{q}}$'s are generalized state vectors (see
\cite{silveirinha_chern_2015}), $\braket{\,\,|\,\,}$ denotes a
weighted inner product and $\partial_\vec{q} \equiv \frac{\partial
}{\partial \vec{q}}$. In the case of a $z$-stratified inhomogeneous
system, the weighted inner product may be defined such that
\begin{equation}
 \braket{\vec{Q}_{n\vec{q}}| \vec{Q}_{n\vec{q}}} = \frac{1}{2} \int_{-d/2}^{d/2} \vec{f}_{n\vec{q}}^\ast \cdot \frac{\partial }{\partial \omega} \left[ \omega \vec{M} \right] \cdot \vec{f}_{n\vec{q}}
 ~dz.
\end{equation}
Furthermore, it can be shown that the numerator of Eq.
\eqref{E:def_Berry_potential} satisfies:
\begin{align}
i \braket{\vec{Q}_{n\vec{q}}|\partial_\vec{q} \vec{Q}_{n\vec{q}}} =
{\mathop{\rm Re}\nolimits} \left\{ { \frac{1}{2} \int_{-d/2}^{d/2}
\vec{f}^\ast_{n\vec{q}} \cdot \frac{\partial }{\partial \omega}
\left[ \omega \vec{M} \right] \cdot i \frac{\partial }{ \partial
\vec{q}}\vec{f}_{n\vec{q}} ~dz } \right\}.
\end{align}
The previous results generalize Eqs. (10) and (14) of
\cite{silveirinha_chern_2015} to the case of $z$-stratified closed
systems. In particular, for the geometry of Fig.
\ref{fig:schematic_structure}, the Berry potential
\eqref{E:def_Berry_potential} is given by
\begin{align}
 \boldsymbol{\mathcal{A}}_{n\vec{q}}= \frac{ \Re{\frac{1}{2} \int_{-d/2}^{d/2} \vec{f}_{n\vec{q}}^\ast \cdot   \partial_\omega(\omega \vec{M}_d)_{\omega_{n\vec{q}}} \cdot i\partial_\vec{q} \vec{f}_{n\vec{q}} dz + \frac{1}{2} \vec{E}_{\text{tan},\vec{q}}^\ast(z=0) \cdot   \partial_\omega(i \boldsymbol{\sigma})_{\omega_{n\vec{q}}} \cdot i \partial_\vec{q} \vec{E}_{\text{tan},\vec{q}}(z=0)} }{ \frac{1}{2}  \int_{-d/2}^{d/2} \vec{f}_{n\vec{q}}^\ast \cdot   \partial_\omega(\omega \vec{M}_d)_{\omega_{n\vec{q}}} \cdot  \vec{f}_{n\vec{q}} dz + \frac{1}{2}  \vec{E}_{\text{tan},\vec{q}}^\ast(z=0) \cdot   \partial_\omega(i \boldsymbol{\sigma})_{\omega_{n\vec{q}}} \cdot  \vec{E}_{\text{tan},\vec{q}}(z=0)  }
\end{align}
where  $\vec{E}_{\text{tan},\vec{q}}=\vec{1}_t  \cdot
\vec{E}_{n\vec{q}}$ is the part of the electric field tangential to
the conductivity sheet and $\partial_\omega \equiv \frac{\partial
}{\partial \omega}$.

As explained in \cite{silveirinha_chern_2015}, for systems invariant
under rotations along the $z$-axis the Chern number of a given band
is simply
\begin{align}
\mathcal{C}_n &= \lim_{q\to \infty} \left(  q \mathcal{A}_{n\vec{q},\varphi}   \right) - \lim_{q\to 0^+} \left(  q \mathcal{A}_{n\vec{q},\varphi}   \right)
\end{align}
where
$\mathcal{A}_{n\vec{q},\varphi}=\boldsymbol{\mathcal{A}}_{n\vec{q}}
\cdot \hat{\boldsymbol{\varphi}}$ and $\hat{\boldsymbol{\varphi}}$
is the azimuthal unit vector in a system of polar coordinates. Even
though the 2D wave vector space is unbounded, for a nondispersive
dielectric host and for the model \eqref{E:QSconductivity} the Chern
number is necessarily an integer because for any band one has
${\omega _{n{\bf{q}}}} \to \infty$ in the ${\bf{q}} \to \infty$
limit, and the response becomes reciprocal when $\omega \to \infty$
\cite{silveirinha_chern_2015}.
%


%

\begin{thebibliography}{56}%
\makeatletter
\providecommand \@ifxundefined [1]{%
 \@ifx{#1\undefined}
}%
\providecommand \@ifnum [1]{%
 \ifnum #1\expandafter \@firstoftwo
 \else \expandafter \@secondoftwo
 \fi
}%
\providecommand \@ifx [1]{%
 \ifx #1\expandafter \@firstoftwo
 \else \expandafter \@secondoftwo
 \fi
}%
\providecommand \natexlab [1]{#1}%
\providecommand \enquote  [1]{``#1''}%
\providecommand \bibnamefont  [1]{#1}%
\providecommand \bibfnamefont [1]{#1}%
\providecommand \citenamefont [1]{#1}%
\providecommand \href@noop [0]{\@secondoftwo}%
\providecommand \href [0]{\begingroup \@sanitize@url \@href}%
\providecommand \@href[1]{\@@startlink{#1}\@@href}%
\providecommand \@@href[1]{\endgroup#1\@@endlink}%
\providecommand \@sanitize@url [0]{\catcode `\\12\catcode `\$12\catcode
  `\&12\catcode `\#12\catcode `\^12\catcode `\_12\catcode `\%12\relax}%
\providecommand \@@startlink[1]{}%
\providecommand \@@endlink[0]{}%
\providecommand \url  [0]{\begingroup\@sanitize@url \@url }%
\providecommand \@url [1]{\endgroup\@href {#1}{\urlprefix }}%
\providecommand \urlprefix  [0]{URL }%
\providecommand \Eprint [0]{\href }%
\providecommand \doibase [0]{http://dx.doi.org/}%
\providecommand \selectlanguage [0]{\@gobble}%
\providecommand \bibinfo  [0]{\@secondoftwo}%
\providecommand \bibfield  [0]{\@secondoftwo}%
\providecommand \translation [1]{[#1]}%
\providecommand \BibitemOpen [0]{}%
\providecommand \bibitemStop [0]{}%
\providecommand \bibitemNoStop [0]{.\EOS\space}%
\providecommand \EOS [0]{\spacefactor3000\relax}%
\providecommand \BibitemShut  [1]{\csname bibitem#1\endcsname}%
\let\auto@bib@innerbib\@empty
\bibitem [{\citenamefont {Hasan}\ and\ \citenamefont
  {Kane}(2010)}]{hasan_textitcolloquium_2010}%
  \BibitemOpen
  \bibfield  {author} {\bibinfo {author} {\bibfnamefont {M.~Z.}\ \bibnamefont
  {Hasan}}\ and\ \bibinfo {author} {\bibfnamefont {C.~L.}\ \bibnamefont
  {Kane}},\ }\bibfield  {title} {\enquote {\bibinfo {title}
  {{\textit{Colloquium}} : {Topological} insulators},}\ }\href {\doibase
  10.1103/RevModPhys.82.3045} {\bibfield  {journal} {\bibinfo  {journal} {Rev.
  Mod. Phys.}\ }\textbf {\bibinfo {volume} {82}},\ \bibinfo {pages}
  {3045--3067} (\bibinfo {year} {2010})}\BibitemShut {NoStop}%
\bibitem [{\citenamefont {Shen}(2012)}]{shen_topological_2012}%
  \BibitemOpen
  \bibfield  {author} {\bibinfo {author} {\bibfnamefont {S.-Q.}\ \bibnamefont
  {Shen}},\ }\href@noop {} {\emph {\bibinfo {title} {Topological
  {Insulators}}}},\ \bibinfo {series} {Springer {Series} in {Solid}-{State}
  {Sciences}}, Vol.\ \bibinfo {volume} {174}\ (\bibinfo  {publisher}
  {Springer},\ \bibinfo {address} {Berlin, Heidelberg},\ \bibinfo {year}
  {2012})\BibitemShut {NoStop}%
\bibitem [{\citenamefont {Lu}\ \emph {et~al.}(2014)\citenamefont {Lu},
  \citenamefont {Joannopoulos},\ and\ \citenamefont
  {Solja\v{c}i\'c}}]{lu_topological_2014}%
  \BibitemOpen
  \bibfield  {author} {\bibinfo {author} {\bibfnamefont {L.}~\bibnamefont
  {Lu}}, \bibinfo {author} {\bibfnamefont {J.~D.}\ \bibnamefont
  {Joannopoulos}}, \ and\ \bibinfo {author} {\bibfnamefont {M.}~\bibnamefont
  {Solja\v{c}i\'c}},\ }\bibfield  {title} {\enquote {\bibinfo {title}
  {Topological photonics},}\ }\href {\doibase 10.1038/nphoton.2014.248}
  {\bibfield  {journal} {\bibinfo  {journal} {Nat. Photonics}\ }\textbf
  {\bibinfo {volume} {8}},\ \bibinfo {pages} {821--829} (\bibinfo {year}
  {2014})}\BibitemShut {NoStop}%
\bibitem [{\citenamefont {Lu}\ \emph {et~al.}(2016)\citenamefont {Lu},
  \citenamefont {Joannopoulos},\ and\ \citenamefont
  {Solja\v{c}i\'c}}]{lu_topological_2016}%
  \BibitemOpen
  \bibfield  {author} {\bibinfo {author} {\bibfnamefont {L.}~\bibnamefont
  {Lu}}, \bibinfo {author} {\bibfnamefont {J.~D.}\ \bibnamefont
  {Joannopoulos}}, \ and\ \bibinfo {author} {\bibfnamefont {M.}~\bibnamefont
  {Solja\v{c}i\'c}},\ }\bibfield  {title} {\enquote {\bibinfo {title}
  {Topological states in photonic systems},}\ }\href {\doibase
  10.1038/nphys3796} {\bibfield  {journal} {\bibinfo  {journal} {Nat. Phys.}\
  }\textbf {\bibinfo {volume} {12}},\ \bibinfo {pages} {626--629} (\bibinfo
  {year} {2016})}\BibitemShut {NoStop}%
\bibitem [{\citenamefont {Haldane}(2017)}]{Haldane_Nobel_2017}%
  \BibitemOpen
  \bibfield  {author} {\bibinfo {author} {\bibfnamefont {F.~D.~M.}\
  \bibnamefont {Haldane}},\ }\bibfield  {title} {\enquote {\bibinfo {title}
  {Nobel lecture: Topological quantum matter},}\ }\href@noop {} {\bibfield
  {journal} {\bibinfo  {journal} {Rev. Mod. Phys.}\ }\textbf {\bibinfo {volume}
  {89}},\ \bibinfo {pages} {040502} (\bibinfo {year} {2017})}\BibitemShut
  {NoStop}%
\bibitem [{\citenamefont {Halperin}(1982)}]{halperin_quantized_1982}%
  \BibitemOpen
  \bibfield  {author} {\bibinfo {author} {\bibfnamefont {B.~I.}\ \bibnamefont
  {Halperin}},\ }\bibfield  {title} {\enquote {\bibinfo {title} {Quantized
  {Hall} conductance, current-carrying edge states, and the existence of
  extended states in a two-dimensional disordered potential},}\ }\href
  {\doibase 10.1103/PhysRevB.25.2185} {\bibfield  {journal} {\bibinfo
  {journal} {Phys. Rev. B}\ }\textbf {\bibinfo {volume} {25}},\ \bibinfo
  {pages} {2185--2190} (\bibinfo {year} {1982})}\BibitemShut {NoStop}%
\bibitem [{\citenamefont {Hatsugai}(1993)}]{hatsugai_chern_1993}%
  \BibitemOpen
  \bibfield  {author} {\bibinfo {author} {\bibfnamefont {Y.}~\bibnamefont
  {Hatsugai}},\ }\bibfield  {title} {\enquote {\bibinfo {title} {Chern number
  and edge states in the integer quantum {Hall} effect},}\ }\href {\doibase
  10.1103/PhysRevLett.71.3697} {\bibfield  {journal} {\bibinfo  {journal}
  {Phys. Rev. Lett.}\ }\textbf {\bibinfo {volume} {71}},\ \bibinfo {pages}
  {3697--3700} (\bibinfo {year} {1993})}\BibitemShut {NoStop}%
\bibitem [{\citenamefont {Kane}\ and\ \citenamefont
  {Mele}(2005)}]{kane_$z_2$_2005}%
  \BibitemOpen
  \bibfield  {author} {\bibinfo {author} {\bibfnamefont {C.~L.}\ \bibnamefont
  {Kane}}\ and\ \bibinfo {author} {\bibfnamefont {E.~J.}\ \bibnamefont
  {Mele}},\ }\bibfield  {title} {\enquote {\bibinfo {title} {{$Z_2$}
  {Topological} {Order} and the {Quantum} {Spin} {Hall} {Effect}},}\ }\href
  {\doibase 10.1103/PhysRevLett.95.146802} {\bibfield  {journal} {\bibinfo
  {journal} {Phys. Rev. Lett.}\ }\textbf {\bibinfo {volume} {95}},\ \bibinfo
  {pages} {146802} (\bibinfo {year} {2005})}\BibitemShut {NoStop}%
\bibitem [{\citenamefont {Raghu}\ and\ \citenamefont
  {Haldane}(2008)}]{raghu_analogs_2008}%
  \BibitemOpen
  \bibfield  {author} {\bibinfo {author} {\bibfnamefont {S.}~\bibnamefont
  {Raghu}}\ and\ \bibinfo {author} {\bibfnamefont {F.~D.~M.}\ \bibnamefont
  {Haldane}},\ }\bibfield  {title} {\enquote {\bibinfo {title} {Analogs of
  quantum-{Hall}-effect edge states in photonic crystals},}\ }\href {\doibase
  10.1103/PhysRevA.78.033834} {\bibfield  {journal} {\bibinfo  {journal} {Phys.
  Rev. A}\ }\textbf {\bibinfo {volume} {78}},\ \bibinfo {pages} {033834}
  (\bibinfo {year} {2008})}\BibitemShut {NoStop}%
\bibitem [{\citenamefont {Silveirinha}(2016)}]{silveirinha_bulk-edge_2016}%
  \BibitemOpen
  \bibfield  {author} {\bibinfo {author} {\bibfnamefont {M.~G.}\ \bibnamefont
  {Silveirinha}},\ }\bibfield  {title} {\enquote {\bibinfo {title} {Bulk-edge
  correspondence for topological photonic continua},}\ }\href {\doibase
  10.1103/PhysRevB.94.205105} {\bibfield  {journal} {\bibinfo  {journal} {Phys.
  Rev. B}\ }\textbf {\bibinfo {volume} {94}},\ \bibinfo {pages} {205105}
  (\bibinfo {year} {2016})}\BibitemShut {NoStop}%
\bibitem [{\citenamefont {Camley}(1987)}]{camley_nonreciprocal_1987}%
  \BibitemOpen
  \bibfield  {author} {\bibinfo {author} {\bibfnamefont {R.~E.}\ \bibnamefont
  {Camley}},\ }\bibfield  {title} {\enquote {\bibinfo {title} {Nonreciprocal
  surface waves},}\ }\href {\doibase 10.1016/0167-5729(87)90006-9} {\bibfield
  {journal} {\bibinfo  {journal} {Surf. Sci. Rep.}\ }\textbf {\bibinfo {volume}
  {7}},\ \bibinfo {pages} {103--187} (\bibinfo {year} {1987})}\BibitemShut
  {NoStop}%
\bibitem [{\citenamefont {Zhukov}\ and\ \citenamefont
  {Raikh}(2000)}]{zhukov_chiral_2000}%
  \BibitemOpen
  \bibfield  {author} {\bibinfo {author} {\bibfnamefont {L.~E.}\ \bibnamefont
  {Zhukov}}\ and\ \bibinfo {author} {\bibfnamefont {M.~E.}\ \bibnamefont
  {Raikh}},\ }\bibfield  {title} {\enquote {\bibinfo {title} {Chiral
  electromagnetic waves at the boundary of optical isomers: {Quantum}
  {Cotton}-{Mouton} effect},}\ }\href {\doibase 10.1103/PhysRevB.61.12842}
  {\bibfield  {journal} {\bibinfo  {journal} {Phys. Rev. B}\ }\textbf {\bibinfo
  {volume} {61}},\ \bibinfo {pages} {12842--12847} (\bibinfo {year}
  {2000})}\BibitemShut {NoStop}%
\bibitem [{\citenamefont {Haldane}\ and\ \citenamefont
  {Raghu}(2008)}]{haldane_possible_2008}%
  \BibitemOpen
  \bibfield  {author} {\bibinfo {author} {\bibfnamefont {F.~D.~M.}\
  \bibnamefont {Haldane}}\ and\ \bibinfo {author} {\bibfnamefont
  {S.}~\bibnamefont {Raghu}},\ }\bibfield  {title} {\enquote {\bibinfo {title}
  {Possible {Realization} of {Directional} {Optical} {Waveguides} in {Photonic}
  {Crystals} with {Broken} {Time}-{Reversal} {Symmetry}},}\ }\href {\doibase
  10.1103/PhysRevLett.100.013904} {\bibfield  {journal} {\bibinfo  {journal}
  {Phys. Rev. Lett.}\ }\textbf {\bibinfo {volume} {100}},\ \bibinfo {pages}
  {013904} (\bibinfo {year} {2008})}\BibitemShut {NoStop}%
\bibitem [{\citenamefont {Yu}\ \emph {et~al.}(2008)\citenamefont {Yu},
  \citenamefont {Veronis}, \citenamefont {Wang},\ and\ \citenamefont
  {Fan}}]{yu_one-way_2008}%
  \BibitemOpen
  \bibfield  {author} {\bibinfo {author} {\bibfnamefont {Z.}~\bibnamefont
  {Yu}}, \bibinfo {author} {\bibfnamefont {G.}~\bibnamefont {Veronis}},
  \bibinfo {author} {\bibfnamefont {Z.}~\bibnamefont {Wang}}, \ and\ \bibinfo
  {author} {\bibfnamefont {S.}~\bibnamefont {Fan}},\ }\bibfield  {title}
  {\enquote {\bibinfo {title} {One-{Way} {Electromagnetic} {Waveguide} {Formed}
  at the {Interface} between a {Plasmonic} {Metal} under a {Static} {Magnetic}
  {Field} and a {Photonic} {Crystal}},}\ }\href {\doibase
  10.1103/PhysRevLett.100.023902} {\bibfield  {journal} {\bibinfo  {journal}
  {Phys. Rev. Lett.}\ }\textbf {\bibinfo {volume} {100}},\ \bibinfo {pages}
  {023902} (\bibinfo {year} {2008})}\BibitemShut {NoStop}%
\bibitem [{\citenamefont {Wang}\ \emph {et~al.}(2009)\citenamefont {Wang},
  \citenamefont {Chong}, \citenamefont {Joannopoulos},\ and\ \citenamefont
  {Solja\v{c}i\'c}}]{wang_observation_2009}%
  \BibitemOpen
  \bibfield  {author} {\bibinfo {author} {\bibfnamefont {Z.}~\bibnamefont
  {Wang}}, \bibinfo {author} {\bibfnamefont {Y.}~\bibnamefont {Chong}},
  \bibinfo {author} {\bibfnamefont {J.~D.}\ \bibnamefont {Joannopoulos}}, \
  and\ \bibinfo {author} {\bibfnamefont {M.}~\bibnamefont {Solja\v{c}i\'c}},\
  }\bibfield  {title} {\enquote {\bibinfo {title} {Observation of
  unidirectional backscattering-immune topological electromagnetic states},}\
  }\href {\doibase 10.1038/nature08293} {\bibfield  {journal} {\bibinfo
  {journal} {Nature}\ }\textbf {\bibinfo {volume} {461}},\ \bibinfo {pages}
  {772--775} (\bibinfo {year} {2009})}\BibitemShut {NoStop}%
\bibitem [{\citenamefont {Ochiai}\ and\ \citenamefont
  {Onoda}(2009)}]{ochiai_photonic_2009}%
  \BibitemOpen
  \bibfield  {author} {\bibinfo {author} {\bibfnamefont {T.}~\bibnamefont
  {Ochiai}}\ and\ \bibinfo {author} {\bibfnamefont {M.}~\bibnamefont {Onoda}},\
  }\bibfield  {title} {\enquote {\bibinfo {title} {Photonic analog of graphene
  model and its extension: {Dirac} cone, symmetry, and edge states},}\ }\href
  {\doibase 10.1103/PhysRevB.80.155103} {\bibfield  {journal} {\bibinfo
  {journal} {Phys. Rev. B}\ }\textbf {\bibinfo {volume} {80}},\ \bibinfo
  {pages} {155103} (\bibinfo {year} {2009})}\BibitemShut {NoStop}%
\bibitem [{\citenamefont {Ao}\ \emph {et~al.}(2009)\citenamefont {Ao},
  \citenamefont {Lin},\ and\ \citenamefont {Chan}}]{ao_one-way_2009}%
  \BibitemOpen
  \bibfield  {author} {\bibinfo {author} {\bibfnamefont {X.}~\bibnamefont
  {Ao}}, \bibinfo {author} {\bibfnamefont {Z.}~\bibnamefont {Lin}}, \ and\
  \bibinfo {author} {\bibfnamefont {C.~T.}\ \bibnamefont {Chan}},\ }\bibfield
  {title} {\enquote {\bibinfo {title} {One-way edge mode in a magneto-optical
  honeycomb photonic crystal},}\ }\href {\doibase 10.1103/PhysRevB.80.033105}
  {\bibfield  {journal} {\bibinfo  {journal} {Phys. Rev. B}\ }\textbf {\bibinfo
  {volume} {80}},\ \bibinfo {pages} {033105} (\bibinfo {year}
  {2009})}\BibitemShut {NoStop}%
\bibitem [{\citenamefont {Poo}\ \emph {et~al.}(2011)\citenamefont {Poo},
  \citenamefont {Wu}, \citenamefont {Lin}, \citenamefont {Yang},\ and\
  \citenamefont {Chan}}]{poo_experimental_2011}%
  \BibitemOpen
  \bibfield  {author} {\bibinfo {author} {\bibfnamefont {Y.}~\bibnamefont
  {Poo}}, \bibinfo {author} {\bibfnamefont {R.-X.}\ \bibnamefont {Wu}},
  \bibinfo {author} {\bibfnamefont {Z.}~\bibnamefont {Lin}}, \bibinfo {author}
  {\bibfnamefont {Y.}~\bibnamefont {Yang}}, \ and\ \bibinfo {author}
  {\bibfnamefont {C.~T.}\ \bibnamefont {Chan}},\ }\bibfield  {title} {\enquote
  {\bibinfo {title} {Experimental {Realization} of {Self}-{Guiding}
  {Unidirectional} {Electromagnetic} {Edge} {States}},}\ }\href {\doibase
  10.1103/PhysRevLett.106.093903} {\bibfield  {journal} {\bibinfo  {journal}
  {Phys. Rev. Lett.}\ }\textbf {\bibinfo {volume} {106}},\ \bibinfo {pages}
  {093903} (\bibinfo {year} {2011})}\BibitemShut {NoStop}%
\bibitem [{\citenamefont {Fang}\ \emph {et~al.}(2012)\citenamefont {Fang},
  \citenamefont {Yu},\ and\ \citenamefont {Fan}}]{fang_realizing_2012}%
  \BibitemOpen
  \bibfield  {author} {\bibinfo {author} {\bibfnamefont {K.}~\bibnamefont
  {Fang}}, \bibinfo {author} {\bibfnamefont {Z.}~\bibnamefont {Yu}}, \ and\
  \bibinfo {author} {\bibfnamefont {S.}~\bibnamefont {Fan}},\ }\bibfield
  {title} {\enquote {\bibinfo {title} {Realizing effective magnetic field for
  photons by controlling the phase of dynamic modulation},}\ }\href {\doibase
  10.1038/nphoton.2012.236} {\bibfield  {journal} {\bibinfo  {journal} {Nat.
  Photonics}\ }\textbf {\bibinfo {volume} {6}},\ \bibinfo {pages} {782--787}
  (\bibinfo {year} {2012})}\BibitemShut {NoStop}%
\bibitem [{\citenamefont {Davoyan}\ and\ \citenamefont
  {Engheta}(2013)}]{davoyan_theory_2013}%
  \BibitemOpen
  \bibfield  {author} {\bibinfo {author} {\bibfnamefont {A.~R.}\ \bibnamefont
  {Davoyan}}\ and\ \bibinfo {author} {\bibfnamefont {N.}~\bibnamefont
  {Engheta}},\ }\bibfield  {title} {\enquote {\bibinfo {title} {Theory of
  {Wave} {Propagation} in {Magnetized} {Near}-{Zero}-{Epsilon} {Metamaterials}:
  {Evidence} for {One}-{Way} {Photonic} {States} and {Magnetically} {Switched}
  {Transparency} and {Opacity}},}\ }\href {\doibase
  10.1103/PhysRevLett.111.257401} {\bibfield  {journal} {\bibinfo  {journal}
  {Phys. Rev. Lett.}\ }\textbf {\bibinfo {volume} {111}},\ \bibinfo {pages}
  {257401} (\bibinfo {year} {2013})}\BibitemShut {NoStop}%
\bibitem [{\citenamefont {Davoyan}\ and\ \citenamefont
  {Engheta}(2014)}]{davoyan_electrically_2014}%
  \BibitemOpen
  \bibfield  {author} {\bibinfo {author} {\bibfnamefont {A.}~\bibnamefont
  {Davoyan}}\ and\ \bibinfo {author} {\bibfnamefont {N.}~\bibnamefont
  {Engheta}},\ }\bibfield  {title} {\enquote {\bibinfo {title} {Electrically
  controlled one-way photon flow in plasmonic nanostructures},}\ }\href
  {\doibase 10.1038/ncomms6250} {\bibfield  {journal} {\bibinfo  {journal}
  {Nat. Commun.}\ }\textbf {\bibinfo {volume} {5}},\ \bibinfo {pages} {5250}
  (\bibinfo {year} {2014})}\BibitemShut {NoStop}%
\bibitem [{\citenamefont {Skirlo}\ \emph {et~al.}(2015)\citenamefont {Skirlo},
  \citenamefont {Lu}, \citenamefont {Igarashi}, \citenamefont {Yan},
  \citenamefont {Joannopoulos},\ and\ \citenamefont
  {Solja\v{c}i\'c}}]{skirlo_experimental_2015}%
  \BibitemOpen
  \bibfield  {author} {\bibinfo {author} {\bibfnamefont {S.~A.}\ \bibnamefont
  {Skirlo}}, \bibinfo {author} {\bibfnamefont {L.}~\bibnamefont {Lu}}, \bibinfo
  {author} {\bibfnamefont {Y.}~\bibnamefont {Igarashi}}, \bibinfo {author}
  {\bibfnamefont {Q.}~\bibnamefont {Yan}}, \bibinfo {author} {\bibfnamefont
  {J.}~\bibnamefont {Joannopoulos}}, \ and\ \bibinfo {author} {\bibfnamefont
  {M.}~\bibnamefont {Solja\v{c}i\'c}},\ }\bibfield  {title} {\enquote {\bibinfo
  {title} {Experimental {Observation} of {Large} {Chern} {Numbers} in
  {Photonic} {Crystals}},}\ }\href {\doibase 10.1103/PhysRevLett.115.253901}
  {\bibfield  {journal} {\bibinfo  {journal} {Phys. Rev. Lett.}\ }\textbf
  {\bibinfo {volume} {115}},\ \bibinfo {pages} {253901} (\bibinfo {year}
  {2015})}\BibitemShut {NoStop}%
\bibitem [{\citenamefont {Abbasi}\ \emph {et~al.}(2015)\citenamefont {Abbasi},
  \citenamefont {Davoyan},\ and\ \citenamefont
  {Engheta}}]{abbasi_one-way_2015}%
  \BibitemOpen
  \bibfield  {author} {\bibinfo {author} {\bibfnamefont {F.}~\bibnamefont
  {Abbasi}}, \bibinfo {author} {\bibfnamefont {A.~R.}\ \bibnamefont {Davoyan}},
  \ and\ \bibinfo {author} {\bibfnamefont {N.}~\bibnamefont {Engheta}},\
  }\bibfield  {title} {\enquote {\bibinfo {title} {One-way surface states due
  to nonreciprocal light-line crossing},}\ }\href {\doibase
  10.1088/1367-2630/17/6/063014} {\bibfield  {journal} {\bibinfo  {journal}
  {New J. Phys.}\ }\textbf {\bibinfo {volume} {17}},\ \bibinfo {pages} {063014}
  (\bibinfo {year} {2015})}\BibitemShut {NoStop}%
\bibitem [{\citenamefont {Minkov}\ and\ \citenamefont
  {Savona}(2016)}]{minkov_haldane_2016}%
  \BibitemOpen
  \bibfield  {author} {\bibinfo {author} {\bibfnamefont {M.}~\bibnamefont
  {Minkov}}\ and\ \bibinfo {author} {\bibfnamefont {V.}~\bibnamefont
  {Savona}},\ }\bibfield  {title} {\enquote {\bibinfo {title} {Haldane quantum
  {Hall} effect for light in a dynamically modulated array of resonators},}\
  }\href {\doibase 10.1364/OPTICA.3.000200} {\bibfield  {journal} {\bibinfo
  {journal} {Optica}\ }\textbf {\bibinfo {volume} {3}},\ \bibinfo {pages}
  {200--206} (\bibinfo {year} {2016})}\BibitemShut {NoStop}%
\bibitem [{\citenamefont {Jin}\ \emph {et~al.}(2016)\citenamefont {Jin},
  \citenamefont {Lu}, \citenamefont {Wang}, \citenamefont {Fang}, \citenamefont
  {Joannopoulos}, \citenamefont {Solja\v{c}i\'c}, \citenamefont {Fu},\ and\
  \citenamefont {Fang}}]{jin_topological_2016}%
  \BibitemOpen
  \bibfield  {author} {\bibinfo {author} {\bibfnamefont {D.}~\bibnamefont
  {Jin}}, \bibinfo {author} {\bibfnamefont {L.}~\bibnamefont {Lu}}, \bibinfo
  {author} {\bibfnamefont {Z.}~\bibnamefont {Wang}}, \bibinfo {author}
  {\bibfnamefont {C.}~\bibnamefont {Fang}}, \bibinfo {author} {\bibfnamefont
  {J.~D.}\ \bibnamefont {Joannopoulos}}, \bibinfo {author} {\bibfnamefont
  {M.}~\bibnamefont {Solja\v{c}i\'c}}, \bibinfo {author} {\bibfnamefont
  {L.}~\bibnamefont {Fu}}, \ and\ \bibinfo {author} {\bibfnamefont {N.~X.}\
  \bibnamefont {Fang}},\ }\bibfield  {title} {\enquote {\bibinfo {title}
  {Topological magnetoplasmon},}\ }\href {\doibase 10.1038/ncomms13486}
  {\bibfield  {journal} {\bibinfo  {journal} {Nat. Commun.}\ }\textbf {\bibinfo
  {volume} {7}},\ \bibinfo {pages} {13486} (\bibinfo {year}
  {2016})}\BibitemShut {NoStop}%
\bibitem [{\citenamefont {He}\ \emph {et~al.}(2016)\citenamefont {He},
  \citenamefont {Sun}, \citenamefont {Liu}, \citenamefont {Lu}, \citenamefont
  {Chen}, \citenamefont {Feng},\ and\ \citenamefont {Chen}}]{he_photonic_2016}%
  \BibitemOpen
  \bibfield  {author} {\bibinfo {author} {\bibfnamefont {C.}~\bibnamefont
  {He}}, \bibinfo {author} {\bibfnamefont {X.-C.}\ \bibnamefont {Sun}},
  \bibinfo {author} {\bibfnamefont {X.-P.}\ \bibnamefont {Liu}}, \bibinfo
  {author} {\bibfnamefont {M.-H.}\ \bibnamefont {Lu}}, \bibinfo {author}
  {\bibfnamefont {Y.}~\bibnamefont {Chen}}, \bibinfo {author} {\bibfnamefont
  {L.}~\bibnamefont {Feng}}, \ and\ \bibinfo {author} {\bibfnamefont {Y.-F.}\
  \bibnamefont {Chen}},\ }\bibfield  {title} {\enquote {\bibinfo {title}
  {Photonic topological insulator with broken time-reversal symmetry},}\ }\href
  {\doibase 10.1073/pnas.1525502113} {\bibfield  {journal} {\bibinfo  {journal}
  {Proceedings of the National Academy of Sciences}\ }\textbf {\bibinfo
  {volume} {113}},\ \bibinfo {pages} {4924--4928} (\bibinfo {year}
  {2016})}\BibitemShut {NoStop}%
\bibitem [{\citenamefont {Jin}\ \emph {et~al.}(2017)\citenamefont {Jin},
  \citenamefont {Christensen}, \citenamefont {Solja\v{c}i\'c}, \citenamefont
  {Fang}, \citenamefont {Lu},\ and\ \citenamefont {Zhang}}]{jin_infrared_2017}%
  \BibitemOpen
  \bibfield  {author} {\bibinfo {author} {\bibfnamefont {D.}~\bibnamefont
  {Jin}}, \bibinfo {author} {\bibfnamefont {T.}~\bibnamefont {Christensen}},
  \bibinfo {author} {\bibfnamefont {M.}~\bibnamefont {Solja\v{c}i\'c}},
  \bibinfo {author} {\bibfnamefont {N.~X.}\ \bibnamefont {Fang}}, \bibinfo
  {author} {\bibfnamefont {L.}~\bibnamefont {Lu}}, \ and\ \bibinfo {author}
  {\bibfnamefont {X.}~\bibnamefont {Zhang}},\ }\bibfield  {title} {\enquote
  {\bibinfo {title} {Infrared {Topological} {Plasmons} in {Graphene}},}\ }\href
  {\doibase 10.1103/PhysRevLett.118.245301} {\bibfield  {journal} {\bibinfo
  {journal} {Phys. Rev. Lett.}\ }\textbf {\bibinfo {volume} {118}},\ \bibinfo
  {pages} {245301} (\bibinfo {year} {2017})}\BibitemShut {NoStop}%
\bibitem [{\citenamefont {Hafezi}\ \emph {et~al.}(2011)\citenamefont {Hafezi},
  \citenamefont {Demler}, \citenamefont {Lukin},\ and\ \citenamefont
  {Taylor}}]{hafezi_robust_2011}%
  \BibitemOpen
  \bibfield  {author} {\bibinfo {author} {\bibfnamefont {M.}~\bibnamefont
  {Hafezi}}, \bibinfo {author} {\bibfnamefont {E.~A.}\ \bibnamefont {Demler}},
  \bibinfo {author} {\bibfnamefont {M.~D.}\ \bibnamefont {Lukin}}, \ and\
  \bibinfo {author} {\bibfnamefont {J.~M.}\ \bibnamefont {Taylor}},\ }\bibfield
   {title} {\enquote {\bibinfo {title} {Robust optical delay lines with
  topological protection},}\ }\href {\doibase 10.1038/nphys2063} {\bibfield
  {journal} {\bibinfo  {journal} {Nat. Phys.}\ }\textbf {\bibinfo {volume}
  {7}},\ \bibinfo {pages} {907--912} (\bibinfo {year} {2011})}\BibitemShut
  {NoStop}%
\bibitem [{\citenamefont {Rechtsman}\ \emph {et~al.}(2013)\citenamefont
  {Rechtsman}, \citenamefont {Zeuner}, \citenamefont {Plotnik}, \citenamefont
  {Lumer}, \citenamefont {Podolsky}, \citenamefont {Dreisow}, \citenamefont
  {Nolte}, \citenamefont {Segev},\ and\ \citenamefont
  {Szameit}}]{rechtsman_photonic_2013}%
  \BibitemOpen
  \bibfield  {author} {\bibinfo {author} {\bibfnamefont {M.~C.}\ \bibnamefont
  {Rechtsman}}, \bibinfo {author} {\bibfnamefont {J.~M.}\ \bibnamefont
  {Zeuner}}, \bibinfo {author} {\bibfnamefont {Y.}~\bibnamefont {Plotnik}},
  \bibinfo {author} {\bibfnamefont {Y.}~\bibnamefont {Lumer}}, \bibinfo
  {author} {\bibfnamefont {D.}~\bibnamefont {Podolsky}}, \bibinfo {author}
  {\bibfnamefont {F.}~\bibnamefont {Dreisow}}, \bibinfo {author} {\bibfnamefont
  {S.}~\bibnamefont {Nolte}}, \bibinfo {author} {\bibfnamefont
  {M.}~\bibnamefont {Segev}}, \ and\ \bibinfo {author} {\bibfnamefont
  {A.}~\bibnamefont {Szameit}},\ }\bibfield  {title} {\enquote {\bibinfo
  {title} {Photonic {Floquet} topological insulators},}\ }\href {\doibase
  10.1038/nature12066} {\bibfield  {journal} {\bibinfo  {journal} {Nature}\
  }\textbf {\bibinfo {volume} {496}},\ \bibinfo {pages} {196--200} (\bibinfo
  {year} {2013})}\BibitemShut {NoStop}%
\bibitem [{\citenamefont {Khanikaev}\ \emph {et~al.}(2013)\citenamefont
  {Khanikaev}, \citenamefont {H.~Mousavi}, \citenamefont {Tse}, \citenamefont
  {Kargarian}, \citenamefont {MacDonald},\ and\ \citenamefont
  {Shvets}}]{khanikaev_photonic_2013}%
  \BibitemOpen
  \bibfield  {author} {\bibinfo {author} {\bibfnamefont {A.~B.}\ \bibnamefont
  {Khanikaev}}, \bibinfo {author} {\bibfnamefont {S.}~\bibnamefont
  {H.~Mousavi}}, \bibinfo {author} {\bibfnamefont {W.-K.}\ \bibnamefont {Tse}},
  \bibinfo {author} {\bibfnamefont {M.}~\bibnamefont {Kargarian}}, \bibinfo
  {author} {\bibfnamefont {A.~H.}\ \bibnamefont {MacDonald}}, \ and\ \bibinfo
  {author} {\bibfnamefont {G.}~\bibnamefont {Shvets}},\ }\bibfield  {title}
  {\enquote {\bibinfo {title} {Photonic topological insulators},}\ }\href
  {\doibase 10.1038/nmat3520} {\bibfield  {journal} {\bibinfo  {journal} {Nat.
  Mat.}\ }\textbf {\bibinfo {volume} {12}},\ \bibinfo {pages} {233--239}
  (\bibinfo {year} {2013})}\BibitemShut {NoStop}%
\bibitem [{\citenamefont {Gao}\ \emph {et~al.}(2015)\citenamefont {Gao},
  \citenamefont {Lawrence}, \citenamefont {Yang}, \citenamefont {Liu},
  \citenamefont {Fang}, \citenamefont {B\'eri}, \citenamefont {Li},\ and\
  \citenamefont {Zhang}}]{gao_topological_2015}%
  \BibitemOpen
  \bibfield  {author} {\bibinfo {author} {\bibfnamefont {W.}~\bibnamefont
  {Gao}}, \bibinfo {author} {\bibfnamefont {M.}~\bibnamefont {Lawrence}},
  \bibinfo {author} {\bibfnamefont {B.}~\bibnamefont {Yang}}, \bibinfo {author}
  {\bibfnamefont {F.}~\bibnamefont {Liu}}, \bibinfo {author} {\bibfnamefont
  {F.}~\bibnamefont {Fang}}, \bibinfo {author} {\bibfnamefont {B.}~\bibnamefont
  {B\'eri}}, \bibinfo {author} {\bibfnamefont {J.}~\bibnamefont {Li}}, \ and\
  \bibinfo {author} {\bibfnamefont {S.}~\bibnamefont {Zhang}},\ }\bibfield
  {title} {\enquote {\bibinfo {title} {Topological {Photonic} {Phase} in
  {Chiral} {Hyperbolic} {Metamaterials}},}\ }\href {\doibase
  10.1103/PhysRevLett.114.037402} {\bibfield  {journal} {\bibinfo  {journal}
  {Phys. Rev. Lett.}\ }\textbf {\bibinfo {volume} {114}},\ \bibinfo {pages}
  {037402} (\bibinfo {year} {2015})}\BibitemShut {NoStop}%
\bibitem [{\citenamefont {Liu}\ and\ \citenamefont
  {Li}(2015)}]{liu_gauge_2015}%
  \BibitemOpen
  \bibfield  {author} {\bibinfo {author} {\bibfnamefont {F.}~\bibnamefont
  {Liu}}\ and\ \bibinfo {author} {\bibfnamefont {J.}~\bibnamefont {Li}},\
  }\bibfield  {title} {\enquote {\bibinfo {title} {Gauge {Field} {Optics} with
  {Anisotropic} {Media}},}\ }\href {\doibase 10.1103/PhysRevLett.114.103902}
  {\bibfield  {journal} {\bibinfo  {journal} {Phys. Rev. Lett.}\ }\textbf
  {\bibinfo {volume} {114}},\ \bibinfo {pages} {103902} (\bibinfo {year}
  {2015})}\BibitemShut {NoStop}%
\bibitem [{\citenamefont {Chen}\ \emph {et~al.}(2015)\citenamefont {Chen},
  \citenamefont {Zhang}, \citenamefont {Dong},\ and\ \citenamefont
  {Chan}}]{chen_symmetry-protected_2015}%
  \BibitemOpen
  \bibfield  {author} {\bibinfo {author} {\bibfnamefont {W.-J.}\ \bibnamefont
  {Chen}}, \bibinfo {author} {\bibfnamefont {Z.-Q.}\ \bibnamefont {Zhang}},
  \bibinfo {author} {\bibfnamefont {J.-W.}\ \bibnamefont {Dong}}, \ and\
  \bibinfo {author} {\bibfnamefont {C.~T.}\ \bibnamefont {Chan}},\ }\bibfield
  {title} {\enquote {\bibinfo {title} {Symmetry-protected transport in a
  pseudospin-polarized waveguide},}\ }\href {\doibase 10.1038/ncomms9183}
  {\bibfield  {journal} {\bibinfo  {journal} {Nat. Commun.}\ }\textbf {\bibinfo
  {volume} {6}},\ \bibinfo {pages} {8183} (\bibinfo {year} {2015})}\BibitemShut
  {NoStop}%
\bibitem [{\citenamefont {Slobozhanyuk}\ \emph {et~al.}(2017)\citenamefont
  {Slobozhanyuk}, \citenamefont {Mousavi}, \citenamefont {Ni}, \citenamefont
  {Smirnova}, \citenamefont {Kivshar},\ and\ \citenamefont
  {Khanikaev}}]{slobozhanyuk_three-dimensional_2017}%
  \BibitemOpen
  \bibfield  {author} {\bibinfo {author} {\bibfnamefont {A.}~\bibnamefont
  {Slobozhanyuk}}, \bibinfo {author} {\bibfnamefont {S.~H.}\ \bibnamefont
  {Mousavi}}, \bibinfo {author} {\bibfnamefont {X.}~\bibnamefont {Ni}},
  \bibinfo {author} {\bibfnamefont {D.}~\bibnamefont {Smirnova}}, \bibinfo
  {author} {\bibfnamefont {Y.~S.}\ \bibnamefont {Kivshar}}, \ and\ \bibinfo
  {author} {\bibfnamefont {A.~B.}\ \bibnamefont {Khanikaev}},\ }\bibfield
  {title} {\enquote {\bibinfo {title} {Three-dimensional all-dielectric
  photonic topological insulator},}\ }\href {\doibase 10.1038/nphoton.2016.253}
  {\bibfield  {journal} {\bibinfo  {journal} {Nat. Photonics}\ }\textbf
  {\bibinfo {volume} {11}},\ \bibinfo {pages} {130--136} (\bibinfo {year}
  {2017})}\BibitemShut {NoStop}%
\bibitem [{\citenamefont {Silveirinha}(2017)}]{silveirinha_PTD_2017}%
  \BibitemOpen
  \bibfield  {author} {\bibinfo {author} {\bibfnamefont {M.~G.}\ \bibnamefont
  {Silveirinha}},\ }\bibfield  {title} {\enquote {\bibinfo {title}
  {{$\mathcal{P}\cdot\mathcal{T}\cdot\mathcal{D}$} symmetry-protected
  scattering anomaly in optics},}\ }\href {\doibase 10.1103/PhysRevB.95.035153}
  {\bibfield  {journal} {\bibinfo  {journal} {Phys. Rev. B}\ }\textbf {\bibinfo
  {volume} {95}},\ \bibinfo {pages} {035153} (\bibinfo {year}
  {2017})}\BibitemShut {NoStop}%
\bibitem [{\citenamefont {Haldane}(1988)}]{haldane_model_1988}%
  \BibitemOpen
  \bibfield  {author} {\bibinfo {author} {\bibfnamefont {F.~D.~M.}\
  \bibnamefont {Haldane}},\ }\bibfield  {title} {\enquote {\bibinfo {title}
  {Model for a {Quantum} {Hall} {Effect} without {Landau} {Levels}:
  {Condensed}-{Matter} {Realization} of the "{Parity} {Anomaly}"},}\ }\href
  {\doibase 10.1103/PhysRevLett.61.2015} {\bibfield  {journal} {\bibinfo
  {journal} {Phys. Rev. Lett.}\ }\textbf {\bibinfo {volume} {61}},\ \bibinfo
  {pages} {2015--2018} (\bibinfo {year} {1988})}\BibitemShut {NoStop}%
\bibitem [{\citenamefont {Kort-Kamp}\ \emph {et~al.}(2015)\citenamefont
  {Kort-Kamp}, \citenamefont {Amorim}, \citenamefont {Bastos}, \citenamefont
  {Pinheiro}, \citenamefont {Rosa}, \citenamefont {Peres},\ and\ \citenamefont
  {Farina}}]{kort-kamp_active_2015}%
  \BibitemOpen
  \bibfield  {author} {\bibinfo {author} {\bibfnamefont {W.~J.~M.}\
  \bibnamefont {Kort-Kamp}}, \bibinfo {author} {\bibfnamefont {B.}~\bibnamefont
  {Amorim}}, \bibinfo {author} {\bibfnamefont {G.}~\bibnamefont {Bastos}},
  \bibinfo {author} {\bibfnamefont {F.~A.}\ \bibnamefont {Pinheiro}}, \bibinfo
  {author} {\bibfnamefont {F.~S.~S.}\ \bibnamefont {Rosa}}, \bibinfo {author}
  {\bibfnamefont {N.~M.~R.}\ \bibnamefont {Peres}}, \ and\ \bibinfo {author}
  {\bibfnamefont {C.}~\bibnamefont {Farina}},\ }\bibfield  {title} {\enquote
  {\bibinfo {title} {Active magneto-optical control of spontaneous emission in
  graphene},}\ }\href {\doibase 10.1103/PhysRevB.92.205415} {\bibfield
  {journal} {\bibinfo  {journal} {Phys. Rev. B}\ }\textbf {\bibinfo {volume}
  {92}},\ \bibinfo {pages} {205415} (\bibinfo {year} {2015})}\BibitemShut
  {NoStop}%
\bibitem [{\citenamefont {Cai}\ \emph {et~al.}(2017)\citenamefont {Cai},
  \citenamefont {Liu}, \citenamefont {Chen}, \citenamefont {Liu}, \citenamefont
  {Shu}, \citenamefont {Luo},\ and\ \citenamefont {Wen}}]{cai_quantized_2017}%
  \BibitemOpen
  \bibfield  {author} {\bibinfo {author} {\bibfnamefont {L.}~\bibnamefont
  {Cai}}, \bibinfo {author} {\bibfnamefont {M.}~\bibnamefont {Liu}}, \bibinfo
  {author} {\bibfnamefont {S.}~\bibnamefont {Chen}}, \bibinfo {author}
  {\bibfnamefont {Y.}~\bibnamefont {Liu}}, \bibinfo {author} {\bibfnamefont
  {W.}~\bibnamefont {Shu}}, \bibinfo {author} {\bibfnamefont {H.}~\bibnamefont
  {Luo}}, \ and\ \bibinfo {author} {\bibfnamefont {S.}~\bibnamefont {Wen}},\
  }\bibfield  {title} {\enquote {\bibinfo {title} {Quantized photonic spin
  {Hall} effect in graphene},}\ }\href {\doibase 10.1103/PhysRevA.95.013809}
  {\bibfield  {journal} {\bibinfo  {journal} {Phys. Rev. A}\ }\textbf {\bibinfo
  {volume} {95}},\ \bibinfo {pages} {013809} (\bibinfo {year}
  {2017})}\BibitemShut {NoStop}%
\bibitem [{\citenamefont {Gibertini}\ \emph {et~al.}(2009)\citenamefont
  {Gibertini}, \citenamefont {Singha}, \citenamefont {Pellegrini},
  \citenamefont {Polini}, \citenamefont {Vignale}, \citenamefont {Pinczuk},
  \citenamefont {Pfeiffer},\ and\ \citenamefont
  {West}}]{gibertini_engineering_2009}%
  \BibitemOpen
  \bibfield  {author} {\bibinfo {author} {\bibfnamefont {M.}~\bibnamefont
  {Gibertini}}, \bibinfo {author} {\bibfnamefont {A.}~\bibnamefont {Singha}},
  \bibinfo {author} {\bibfnamefont {V.}~\bibnamefont {Pellegrini}}, \bibinfo
  {author} {\bibfnamefont {M.}~\bibnamefont {Polini}}, \bibinfo {author}
  {\bibfnamefont {G.}~\bibnamefont {Vignale}}, \bibinfo {author} {\bibfnamefont
  {A.}~\bibnamefont {Pinczuk}}, \bibinfo {author} {\bibfnamefont {L.~N.}\
  \bibnamefont {Pfeiffer}}, \ and\ \bibinfo {author} {\bibfnamefont {K.~W.}\
  \bibnamefont {West}},\ }\bibfield  {title} {\enquote {\bibinfo {title}
  {Engineering artificial graphene in a two-dimensional electron gas},}\ }\href
  {\doibase 10.1103/PhysRevB.79.241406} {\bibfield  {journal} {\bibinfo
  {journal} {Phys. Rev. B}\ }\textbf {\bibinfo {volume} {79}},\ \bibinfo
  {pages} {241406} (\bibinfo {year} {2009})}\BibitemShut {NoStop}%
\bibitem [{\citenamefont {Lanneb\`ere}\ and\ \citenamefont
  {Silveirinha}(2015)}]{lannebere_effective_2015}%
  \BibitemOpen
  \bibfield  {author} {\bibinfo {author} {\bibfnamefont {S.}~\bibnamefont
  {Lanneb\`ere}}\ and\ \bibinfo {author} {\bibfnamefont {M.~G.}\ \bibnamefont
  {Silveirinha}},\ }\bibfield  {title} {\enquote {\bibinfo {title} {Effective
  {Hamiltonian} for electron waves in artificial graphene: {A} first-principles
  derivation},}\ }\href {\doibase 10.1103/PhysRevB.91.045416} {\bibfield
  {journal} {\bibinfo  {journal} {Phys. Rev. B}\ }\textbf {\bibinfo {volume}
  {91}},\ \bibinfo {pages} {045416} (\bibinfo {year} {2015})}\BibitemShut
  {NoStop}%
\bibitem [{\citenamefont {Silveirinha}(2015)}]{silveirinha_chern_2015}%
  \BibitemOpen
  \bibfield  {author} {\bibinfo {author} {\bibfnamefont {M.~G.}\ \bibnamefont
  {Silveirinha}},\ }\bibfield  {title} {\enquote {\bibinfo {title} {Chern
  invariants for continuous media},}\ }\href {\doibase
  10.1103/PhysRevB.92.125153} {\bibfield  {journal} {\bibinfo  {journal} {Phys.
  Rev. B}\ }\textbf {\bibinfo {volume} {92}},\ \bibinfo {pages} {125153}
  (\bibinfo {year} {2015})}\BibitemShut {NoStop}%
\bibitem [{\citenamefont {Silveirinha}\ and\ \citenamefont
  {Engheta}(2012)}]{silveirinha_effective_2012}%
  \BibitemOpen
  \bibfield  {author} {\bibinfo {author} {\bibfnamefont {M.~G.}\ \bibnamefont
  {Silveirinha}}\ and\ \bibinfo {author} {\bibfnamefont {N.}~\bibnamefont
  {Engheta}},\ }\bibfield  {title} {\enquote {\bibinfo {title} {Effective
  medium approach to electron waves: {Graphene} superlattices},}\ }\href
  {\doibase 10.1103/PhysRevB.85.195413} {\bibfield  {journal} {\bibinfo
  {journal} {Phys. Rev. B}\ }\textbf {\bibinfo {volume} {85}},\ \bibinfo
  {pages} {195413} (\bibinfo {year} {2012})}\BibitemShut {NoStop}%
\bibitem [{\citenamefont {Nagaosa}\ \emph {et~al.}(2010)\citenamefont
  {Nagaosa}, \citenamefont {Sinova}, \citenamefont {Onoda}, \citenamefont
  {MacDonald},\ and\ \citenamefont {Ong}}]{Nagaosa}%
  \BibitemOpen
  \bibfield  {author} {\bibinfo {author} {\bibfnamefont {N.}~\bibnamefont
  {Nagaosa}}, \bibinfo {author} {\bibfnamefont {J.}~\bibnamefont {Sinova}},
  \bibinfo {author} {\bibfnamefont {S.}~\bibnamefont {Onoda}}, \bibinfo
  {author} {\bibfnamefont {A.~H.}\ \bibnamefont {MacDonald}}, \ and\ \bibinfo
  {author} {\bibfnamefont {N.~P.}\ \bibnamefont {Ong}},\ }\bibfield  {title}
  {\enquote {\bibinfo {title} {Anomalous hall effect},}\ }\href@noop {}
  {\bibfield  {journal} {\bibinfo  {journal} {Rev. Mod. Phys.}\ }\textbf
  {\bibinfo {volume} {82}},\ \bibinfo {pages} {1539} (\bibinfo {year}
  {2010})}\BibitemShut {NoStop}%
\bibitem [{\citenamefont {Thouless}\ \emph {et~al.}(1982)\citenamefont
  {Thouless}, \citenamefont {Kohmoto}, \citenamefont {Nightingale},\ and\
  \citenamefont {den Nijs}}]{thouless_quantized_1982}%
  \BibitemOpen
  \bibfield  {author} {\bibinfo {author} {\bibfnamefont {D.~J.}\ \bibnamefont
  {Thouless}}, \bibinfo {author} {\bibfnamefont {M.}~\bibnamefont {Kohmoto}},
  \bibinfo {author} {\bibfnamefont {M.~P.}\ \bibnamefont {Nightingale}}, \ and\
  \bibinfo {author} {\bibfnamefont {M.}~\bibnamefont {den Nijs}},\ }\bibfield
  {title} {\enquote {\bibinfo {title} {Quantized {Hall} {Conductance} in a
  {Two}-{Dimensional} {Periodic} {Potential}},}\ }\href {\doibase
  10.1103/PhysRevLett.49.405} {\bibfield  {journal} {\bibinfo  {journal} {Phys.
  Rev. Lett.}\ }\textbf {\bibinfo {volume} {49}},\ \bibinfo {pages} {405--408}
  (\bibinfo {year} {1982})}\BibitemShut {NoStop}%
\bibitem [{\citenamefont {Kubo}(1966)}]{Kubo}%
  \BibitemOpen
  \bibfield  {author} {\bibinfo {author} {\bibfnamefont {R.}~\bibnamefont
  {Kubo}},\ }\bibfield  {title} {\enquote {\bibinfo {title} {The
  fluctuation-dissipation theorem},}\ }\href@noop {} {\bibfield  {journal}
  {\bibinfo  {journal} {Rep. Prog. Phys.}\ }\textbf {\bibinfo {volume} {29}},\
  \bibinfo {pages} {255} (\bibinfo {year} {1966})}\BibitemShut {NoStop}%
\bibitem [{\citenamefont {Mikhailov}\ and\ \citenamefont
  {Ziegler}(2007)}]{mikhailov_new_2007}%
  \BibitemOpen
  \bibfield  {author} {\bibinfo {author} {\bibfnamefont {S.~A.}\ \bibnamefont
  {Mikhailov}}\ and\ \bibinfo {author} {\bibfnamefont {K.}~\bibnamefont
  {Ziegler}},\ }\bibfield  {title} {\enquote {\bibinfo {title} {New
  {Electromagnetic} {Mode} in {Graphene}},}\ }\href {\doibase
  10.1103/PhysRevLett.99.016803} {\bibfield  {journal} {\bibinfo  {journal}
  {Phys. Rev. Lett.}\ }\textbf {\bibinfo {volume} {99}},\ \bibinfo {pages}
  {016803} (\bibinfo {year} {2007})}\BibitemShut {NoStop}%
\bibitem [{\citenamefont {Goncalves}\ and\ \citenamefont
  {Peres}(2016)}]{goncalves_introduction_2016}%
  \BibitemOpen
  \bibfield  {author} {\bibinfo {author} {\bibfnamefont {P.~A.~D.}\
  \bibnamefont {Goncalves}}\ and\ \bibinfo {author} {\bibfnamefont {N.~M.~R.}\
  \bibnamefont {Peres}},\ }\href@noop {} {\emph {\bibinfo {title} {An
  {Introduction} to {Graphene} {Plasmonics}}}}\ (\bibinfo  {publisher} {World
  Scientific Publishing Co Pte Ltd},\ \bibinfo {address} {New Jersey},\
  \bibinfo {year} {2016})\BibitemShut {NoStop}%
\bibitem [{\citenamefont {Allen}(2006)}]{Allen}%
  \BibitemOpen
  \bibfield  {author} {\bibinfo {author} {\bibfnamefont {P.~B.}\ \bibnamefont
  {Allen}},\ }\href@noop {} {\emph {\bibinfo {title} {Electron transport}}}\
  (\bibinfo  {publisher} {book chapter in ``Conceptual Properties of Materials,
  A Standard Model for Ground- and Excited-State Properties'', edited by S. G.
  Louie and M. L. Cohen, Elsevier},\ \bibinfo {address} {Amsterdam},\ \bibinfo
  {year} {2006})\BibitemShut {NoStop}%
\bibitem [{\citenamefont {Bittencourt}(2010)}]{Bittencourt}%
  \BibitemOpen
  \bibfield  {author} {\bibinfo {author} {\bibfnamefont {J.~A.}\ \bibnamefont
  {Bittencourt}},\ }\href@noop {} {\emph {\bibinfo {title} {Fundamentals of
  Plasma Physics, 3rd Ed.}}}\ (\bibinfo  {publisher} {Springer-Verlag},\
  \bibinfo {address} {New York},\ \bibinfo {year} {2010})\BibitemShut {NoStop}%
\bibitem [{\citenamefont {Chiu}\ and\ \citenamefont
  {Quinn}(1974)}]{chiu_plasma_1974}%
  \BibitemOpen
  \bibfield  {author} {\bibinfo {author} {\bibfnamefont {K.~W.}\ \bibnamefont
  {Chiu}}\ and\ \bibinfo {author} {\bibfnamefont {J.~J.}\ \bibnamefont
  {Quinn}},\ }\bibfield  {title} {\enquote {\bibinfo {title} {Plasma
  oscillations of a two-dimensional electron gas in a strong magnetic field},}\
  }\href {\doibase 10.1103/PhysRevB.9.4724} {\bibfield  {journal} {\bibinfo
  {journal} {Phys. Rev. B}\ }\textbf {\bibinfo {volume} {9}},\ \bibinfo {pages}
  {4724--4732} (\bibinfo {year} {1974})}\BibitemShut {NoStop}%
\bibitem [{\citenamefont {Ferreira}\ \emph {et~al.}(2012)\citenamefont
  {Ferreira}, \citenamefont {Peres},\ and\ \citenamefont
  {Castro~Neto}}]{ferreira_confined_2012}%
  \BibitemOpen
  \bibfield  {author} {\bibinfo {author} {\bibfnamefont {A.}~\bibnamefont
  {Ferreira}}, \bibinfo {author} {\bibfnamefont {N.~M.~R.}\ \bibnamefont
  {Peres}}, \ and\ \bibinfo {author} {\bibfnamefont {A.~H.}\ \bibnamefont
  {Castro~Neto}},\ }\bibfield  {title} {\enquote {\bibinfo {title} {Confined
  magneto-optical waves in graphene},}\ }\href {\doibase
  10.1103/PhysRevB.85.205426} {\bibfield  {journal} {\bibinfo  {journal} {Phys.
  Rev. B}\ }\textbf {\bibinfo {volume} {85}},\ \bibinfo {pages} {205426}
  (\bibinfo {year} {2012})}\BibitemShut {NoStop}%
\bibitem [{\citenamefont {Avron}\ \emph {et~al.}(1983)\citenamefont {Avron},
  \citenamefont {Seiler},\ and\ \citenamefont {Simon}}]{avron_homotopy_1983}%
  \BibitemOpen
  \bibfield  {author} {\bibinfo {author} {\bibfnamefont {J.~E.}\ \bibnamefont
  {Avron}}, \bibinfo {author} {\bibfnamefont {R.}~\bibnamefont {Seiler}}, \
  and\ \bibinfo {author} {\bibfnamefont {B.}~\bibnamefont {Simon}},\ }\bibfield
   {title} {\enquote {\bibinfo {title} {Homotopy and {Quantization} in
  {Condensed} {Matter} {Physics}},}\ }\href {\doibase
  10.1103/PhysRevLett.51.51} {\bibfield  {journal} {\bibinfo  {journal} {Phys.
  Rev. Lett.}\ }\textbf {\bibinfo {volume} {51}},\ \bibinfo {pages} {51--53}
  (\bibinfo {year} {1983})}\BibitemShut {NoStop}%
\bibitem [{CST()}]{CST}%
  \BibitemOpen
  CST,\ \href@noop {} {\enquote {\bibinfo {title} {{GmbH 2017 CST Microwave
  Studio}},}\ }\bibinfo {howpublished} {\url{http://www.cst.com}} (\bibinfo
  {year} {2017})\BibitemShut {NoStop}%
\bibitem [{Supplemental Material with the time animation of Fig. 6a, showing
  the propagation of the unidirectional edge state along the
  interface()}]{Sup_material}%
  \BibitemOpen
  Supplemental Material with the time animation of Fig. 6a, showing the
  propagation of the unidirectional edge state along the interface,\ \href@noop
  {} {}\BibitemShut {NoStop}%
\bibitem [{\citenamefont {Gangaraj}\ \emph {et~al.}(2016)\citenamefont
  {Gangaraj}, \citenamefont {Nemilentsau},\ and\ \citenamefont
  {Hanson}}]{gangaraj_effects_2016}%
  \BibitemOpen
  \bibfield  {author} {\bibinfo {author} {\bibfnamefont {S.~A.~H.}\
  \bibnamefont {Gangaraj}}, \bibinfo {author} {\bibfnamefont {A.}~\bibnamefont
  {Nemilentsau}}, \ and\ \bibinfo {author} {\bibfnamefont {G.~W.}\ \bibnamefont
  {Hanson}},\ }\bibfield  {title} {\enquote {\bibinfo {title} {The effects of
  three-dimensional defects on one-way surface plasmon propagation for photonic
  topological insulators comprised of continuum media},}\ }\href {\doibase
  10.1038/srep30055} {\bibfield  {journal} {\bibinfo  {journal} {Sci. Rep.}\
  }\textbf {\bibinfo {volume} {6}},\ \bibinfo {pages} {30055} (\bibinfo {year}
  {2016})}\BibitemShut {NoStop}%
\bibitem [{\citenamefont {Silveirinha}(to appear; available online in
  arxiv.org/abs/1712.04272)}]{SilvDisp}%
  \BibitemOpen
  \bibfield  {author} {\bibinfo {author} {\bibfnamefont {M.~G.}\ \bibnamefont
  {Silveirinha}},\ }\href@noop {} {\emph {\bibinfo {title} {Modal expansions in
  dispersive material systems with application to quantum optics and
  topological photonics}}}\ (\bibinfo  {publisher} {book chapter in ``Advances
  in Mathematical Methods for Electromagnetics"", edited by Paul Smith and
  Kazuya Kobayashi, published by IET},\ \bibinfo {year} {to appear; available
  online in arxiv.org/abs/1712.04272})\BibitemShut {NoStop}%
\end{thebibliography}

\end{document}